\documentclass[final,1p,times,twocolumn]{elsarticle}

\usepackage{amsmath,amssymb}
\usepackage{booktabs}
\usepackage[ruled,vlined]{algorithm2e}
\usepackage{array}
\usepackage[colorlinks=true, allcolors=blue]{hyperref}
\usepackage{xcolor}
\usepackage{graphicx}
\usepackage{multirow}
\usepackage{enumitem}
\usepackage{lineno}
\usepackage{subcaption}
\usepackage{siunitx}
\graphicspath{{figures/}}
\usepackage{soul}
\usepackage[utf8]{inputenc}
\usepackage{bm}
\journal{International Journal of Mechanical Sciences}
\begin{document}

\begin{frontmatter}

\title{Amortized Posteriors for Estimation of Material Constitutive Parameters 
from Multimodal Measurements on Small Punch Tests}
\author[gt]{Mohammad Ali Seyed Mahmoud\corref{cor1}}
\ead{m.seyedmahmoud@gatech.edu}
\author[mult]{Aditya Venkatraman}
\author[gt]{Raj Mahat}
\author[gt,cse]{Samantha Mitra}
\author[gt,cse,mult]{Surya R. Kalidindi\corref{cor1}}
\ead{surya.kalidindi@me.gatech.edu}
\cortext[cor1]{Corresponding authors}

\address[gt]{George W. Woodruff School of Mechanical Engineering,
            Georgia Institute of Technology, Atlanta, GA, USA}
\address[cse]{School of Computational Science and Engineering,
             Georgia Institute of Technology, Atlanta, GA, USA}
\address[mult]{Multiscale Technologies Inc, Atlanta, GA, USA}

\begin{abstract}
Bayesian calibration of material constitutive parameters from multimodal 
mechanical test data is
often limited by the need to specify a joint likelihood across measurement
modalities that differ in dimensionality, noise structure, and physical
units.
The resulting posteriors are often broad or strongly correlated, causing
standard Markov Chain Monte Carlo (MCMC) samplers to mix poorly. Here, 
we present an amortized, likelihood-free framework that combines 
Gaussian process (GP) surrogates with Conditional Flow Matching (CFM) to 
learn conditional posteriors over constitutive parameters directly from 
synthetic multimodal parameter--observation pairs, avoiding hand-crafted likelihoods
and repeated MCMC sampling. Once trained, the GP--CFM model generates posterior
samples for each new specimen at negligible cost. The utility of this novel approach is
demonstrated in this paper by estimating the values of Young's modulus
and yield strength from the early portion of the force--displacement ($F$--$D$) curve and a 
Digital Image Correlation (DIC)-based displacement field measured in a Small Punch Test (SPT). 
It is observed that the $F$--$D$ data alone produce broad posteriors, consistent with 
limited parameter discrimination in the global response. Adding the DIC-measured 
displacement field was seen to contract the posteriors and shift them
towards the independently measured tensile reference
values. This work establishes a robust likelihood-free framework for 
the inference of material constitutive parameters from multimodal data, 
demonstrated through SPT--DIC integration.
\end{abstract}


\end{frontmatter}


\section{Introduction}
\label{sec:intro}

High-throughput, miniaturized, and non-standard mechanical testing
approaches are gaining increasing attention for material constitutive
calibration because they reduce material volume requirements, testing
time, and cost relative to conventional uniaxial experiments
\cite{manahan1981development,garcia2014review,Mahmoud2023Materials,ASTM_E3205_20}.
Beyond efficiency, many of these tests impose complex, spatially
inhomogeneous loading paths that activate multiple deformation mechanisms
simultaneously, and when combined with modern full-field sensing
technologies, they can produce rich multimodal data comprising both global
mechanical responses and spatially resolved deformation fields
\cite{sutton2009image,pan2009dic}. However, the same
mechanical complexity that makes these tests informative also makes their
interpretation difficult; the measured global response (for example, a 
force-displacement curve or the energy absorbed by the specimen) typically couples
contributions from several constitutive material parameters so that different
parameter combinations can produce nearly indistinguishable global
responses \cite{tarantola2005inverse,kaipio2005statistical}. 
Reliable estimation of material constitutive parameters from such data therefore
requires analysis methods that can handle multimodal measurements 
and account for the limited identifiability inherent in mechanically complex tests.

The Small Punch Test (SPT) exemplifies both the promise and the
difficulty of this setting. It is used for mechanical characterization
when material volume precludes conventional tensile specimens, as in
irradiated materials, weld heat-affected zones, and thin
coatings~\cite{manahan1981development,garcia2014review,leclerc2021spt,ASTM_E3205_20}.
During loading, the punch drives a clamped miniature specimen through a
sequence of elastic bending, plastic deformation, contact evolution, and
membrane stretching, producing a spatially inhomogeneous stress state
that limits the reliability of simple empirical correlations or direct
curve fitting for joint parameter
identification~\cite{campitelli2004assessment,abendroth2006identification}.
A particular difficulty is that, unlike a uniaxial tensile test, the SPT
force--displacement curve does not provide an easily identified elastic-plastic
transition. Chica et al.~\cite{chica2017elastic} showed through
finite element (FE) analysis that the early portion of the SPT curve contains
both elastic bending of the specimen and local plastic indentation beneath
the punch. As a result, the initial slope of the SPT curve depends on both
elastic and plastic material properties, rather than on Young's modulus
alone. Not surprisingly, many prior studies have fixed the elastic
modulus, typically from an independent tensile test, instead of estimating
$E$ (Young's modulus) and $\sigma_y$ (yield strength) jointly from the SPT measurements
alone~\cite{chica2017elastic,campitelli2004assessment}.
This specific  identifiability problem has motivated the present work and is described in detail next using results of
FE simulations. 

\begin{figure}[!t]
\centering
\begin{subfigure}[b]{\columnwidth}
    \centering
    \includegraphics[width=0.72\columnwidth]{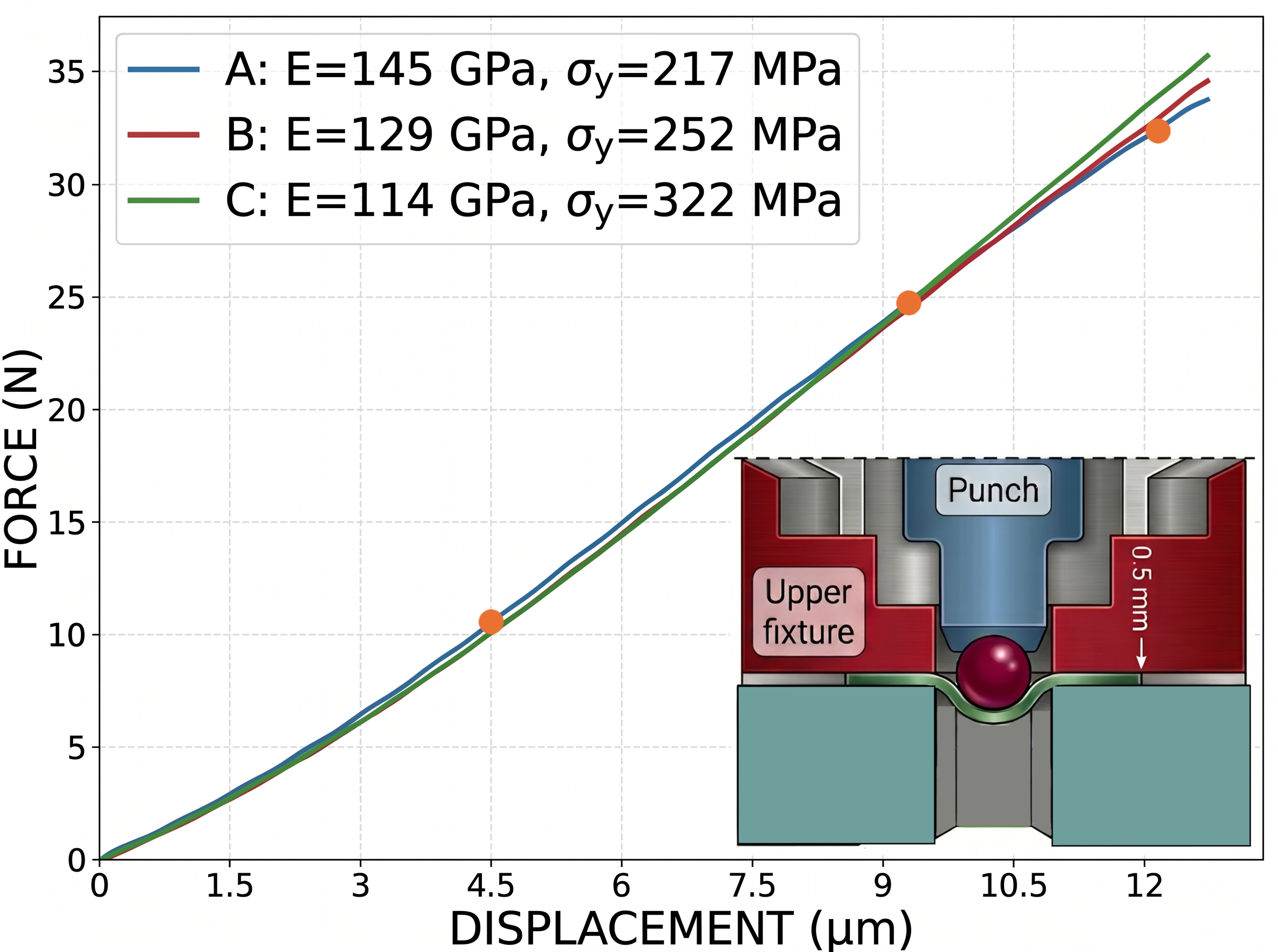}
    \caption{}
    \label{fig:intro_coupling_fd}
\end{subfigure}
\vspace{0.5em}
\begin{subfigure}[b]{\columnwidth}
    \centering
    \includegraphics[width=\columnwidth]{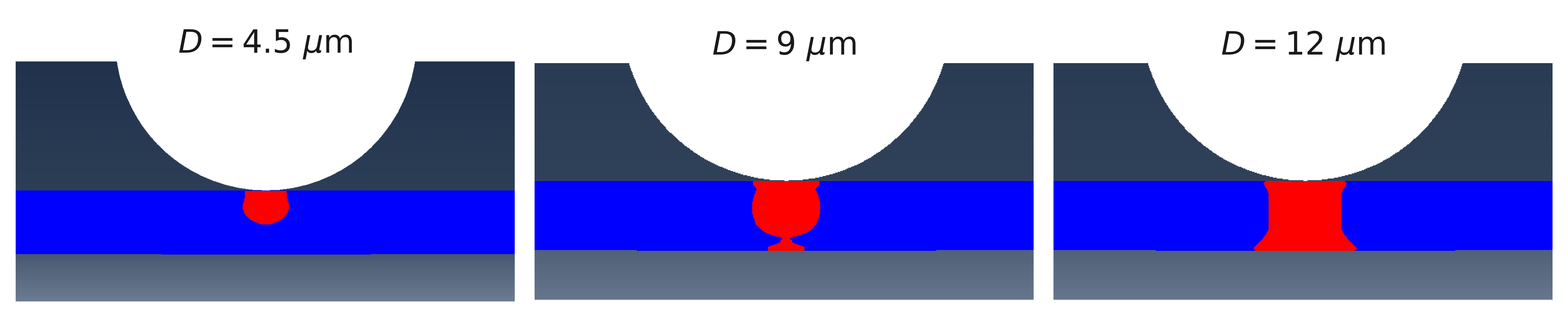}
    \caption{}
    \label{fig:intro_coupling_peeq}
\end{subfigure}
\caption{Parameter coupling in the early SPT response. (a) Simulated early
force--displacement responses for three combinations of Young's modulus $E$
and yield strength $\sigma_y$; markers denote punch displacements
$D = 4.5$, $9$, and $12~\mu\mathrm{m}$ on curve~A, and the inset shows
the FE model of the SPT configuration. (b) Regions satisfying the von Mises
yield condition (red) beneath the punch at the three marked displacements.}
\label{fig:intro_coupling}
\end{figure}

Figure~\ref{fig:intro_coupling}a demonstrates that different
combinations of values for $E$ and $\sigma_y$ can produce nearly indistinguishable
early force--displacement responses. Figure~\ref{fig:intro_coupling}b shows the
regions satisfying the von Mises yield condition beneath the punch at three
displacements along one of these curves. Yielding initiates at very small
punch displacement (see the contour corresponding to $D=4.5~\mu\mathrm{m}$).
The hourglass-shaped plastic yield region results from the coalescence of a
contact-driven plastic zone beneath the punch and a tensile bending-driven
plastic zone at the lower surface. Before coalescence, an elastic core remains
through part of the specimen thickness, and the response is governed
predominantly by elastic bending and the onset of yielding. Once a continuous
plastic zone forms through the thickness, deformation progressively
transitions toward generalized plastic bending and membrane stretching.
Plastic strain then starts accumulating rapidly, making the
mechanical response increasingly sensitive to the hardening law. However, before 
the coalescence of the continuous through-thickness plastic zone, the 
force-displacement response depends mainly on the values of $E$ and $\sigma_y$ 
of the material, and is relatively insensitive to the hardening parameters 
because of the limited plasticity. Therefore, there exists an
\textit{identifiability challenge} for the simultaneous estimation of both 
$E$ and $\sigma_y$ from this early portion of the force--displacement curve. 
This, in turn, motivates the incorporation of additional measurement modalities. 
Specifically, recent studies have demonstrated the
feasibility of Digital Image Correlation
(DIC)-based measurements of displacement and strain fields on the bottom surface
of the SPT
sample~\cite{vijayanand2020novel}. These advances have set the stage
for the present work by
raising the possibility of joint estimation of both $E$ and $\sigma_y$ from the
measured force--displacement
curves in combination with the DIC measurements in SPT, to overcome the 
identifiability challenge. 

To the best of our knowledge, this study is the first attempt to recover both
$E$ and $\sigma_y$ from a DIC-instrumented SPT. The constitutive description
is deliberately restricted to an elastic--perfectly plastic model and to the
early response preceding coalescence of the through-thickness plastic zone.
This controlled setting isolates the joint identifiability problem and allows
the inferred parameters to be evaluated against independent tensile
measurements. Extension to hardening parameters requires later-stage response
data and is outside the present scope.

Existing bottom-surface SPT--DIC measurements have commonly relied on
mirror-based optical access, which can restrict the effective stereo angle
and introduce additional optical-path and reflection-plane distortions
\cite{vijayanand2020novel}. The direct-view stereoscopic DIC configuration
developed in this work avoids these limitations and provides the
bottom-surface displacement measurements used for parameter inference.

Bayesian inference is the standard probabilistic framework for
constitutive calibration because it naturally quantifies parameter
uncertainty, reveals correlations, and exposes
non-identifiability~\cite{kaipio2005statistical,venkatraman2022bayesian}.
When the calibration data come from multiple measurement modalities, denoted here generically as $( \mathbf{s}_1, \mathbf{s}_2)$, the posterior over the material constitutive parameters ($\bm{\theta}$)  can be expressed as

\begin{equation}
    p(\bm{\theta} \mid \mathbf{s}_1, \mathbf{s}_2)
    \;\propto\;
    p(\mathbf{s}_1, \mathbf{s}_2 \mid \bm{\theta})\,p(\bm{\theta}),
    \label{eq:bayes_joint}
\end{equation}
where 
$p(\mathbf{s}_1, \mathbf{s}_2 \mid \bm{\theta})$ represents a joint likelihood that is difficult to
compute. For a Gaussian residual model this takes the form
\begin{equation}
    p(\mathbf{s}_1, \mathbf{s}_2 \mid \bm{\theta})
    \;=\;
    \mathcal{N}\!\!\left(
    \begin{bmatrix} \mathbf{s}_1 \\ \mathbf{s}_2 \end{bmatrix}
    \;\Bigg|\;
    \begin{bmatrix} \bm{\mu}_1(\bm{\theta}) \\
    \bm{\mu}_2(\bm{\theta}) \end{bmatrix},\;
    \begin{bmatrix}
      \bm{\Sigma}_{11} & \bm{\Sigma}_{12} \\
      \bm{\Sigma}_{21} & \bm{\Sigma}_{22}
    \end{bmatrix}
    \right),
    \label{eq:joint_gaussian}
\end{equation}
where $\bm{\mu}_1(\bm{\theta})$ and $\bm{\mu}_2(\bm{\theta})$ are the
forward-model predictions for each modality, $\bm{\Sigma}_{11}$ and
$\bm{\Sigma}_{22}$ are the within-modality residual covariances, and
$\bm{\Sigma}_{12}$ is the cross-covariance between residuals from the
two modalities. The within-modality terms can, in principle, be estimated
from experimental replicates. The cross-covariance $\bm{\Sigma}_{12}$
is difficult to estimate reliably, because the two modalities are likely to differ
in physical units, dimensionality, noise mechanisms, and spatial
structure; no existing straightforward replication strategy isolates this term
from forward-model discrepancy. Setting
$\bm{\Sigma}_{12} = \mathbf{0}$ implies a significant approximation of the likelihood as $
    p(\mathbf{s}_1, \mathbf{s}_2 \mid \bm{\theta})
    \;\approx\;
    p(\mathbf{s}_1 \mid \bm{\theta})\,
    p(\mathbf{s}_2 \mid \bm{\theta})$, which can lead to
overconfident posteriors. Moreover, even when a likelihood model is accepted, the
resulting posterior can be strongly correlated or weakly identifiable,
causing standard Markov Chain Monte Carlo (MCMC) samplers to mix
poorly~\cite{girolami2011riemann,beskos2017geometric,neal2011mcmc,venkatraman2025bayesian}.
Bayesian calibration combining force--displacement and full-field DIC
data has been demonstrated for other test geometries using
MCMC~\cite{ricciardi2025icc,seyedmahmoud2026sequential}, but the
per-specimen cost of MCMC and the need for an explicit likelihood limit
scalability. Together, the difficulty of specifying a reliable joint likelihood across
measurement modalities and the cost of sampling correlated posteriors motivate an
inference approach that avoids explicit likelihood construction. Sequential updating
strategies, in which the posterior from one modality
serves as the prior for the next, introduce a further complication: our
earlier work found that the resulting posterior can depend on the order in
which the modalities are incorporated~\cite{seyedmahmoud2026sequential}. The
approach developed here avoids this undesired dependence by incorporating both modalities simultaneously.

In this study, we develop a computational--experimental framework for
learning likelihood-free conditional posteriors over constitutive parameters
from SPT--DIC measurements by combining Gaussian process (GP)
surrogates~\cite{rasmussen2006gp,castillo2019bayesian} with Conditional Flow
Matching (CFM)~\cite{lipman2023flow,albergo2023stochastic}. The conditional
posterior is learned directly from synthetic parameter--observation pairs
generated by the forward
model~\cite{cranmer2020frontier,papamakarios2021normalizing}, avoiding
explicit joint-likelihood construction; once trained, the estimator is
amortized, producing posterior samples for each new specimen at low marginal
cost. As already mentioned, we focus specifically on the joint
identification of Young's modulus $E$ and yield strength $\sigma_y$. Towards this
goal, we employ an elastic--perfectly plastic (EPP) constitutive model, and
demonstrate the benefits of our new approach on SPT measurements performed on an
AA6111-T4 aluminum alloy. This study
makes three distinct contributions:
\begin{enumerate}[label=(\roman*)]

  \item A GP--CFM amortized inference framework for constitutive
        calibration that learns conditional posteriors from synthetic
        parameter--observation pairs, avoiding explicit joint-likelihood
        construction and repeated MCMC sampling at inference time.

  \item A direct-view stereoscopic DIC protocol for quantitative
        out-of-plane displacement measurement on the SPT specimen bottom
        surface, avoiding the stereo-angle, distortion, and resolution
        penalties associated with mirror-based arrangements.

  \item A demonstration that bottom-surface DIC displacement information improves joint posterior inference of
        $E$ and $\sigma_y$ relative to force--displacement data alone, producing
        contracted posteriors consistent with independently measured tensile
        reference properties.
\end{enumerate}

\section{Experiments}
\label{sec:experiment}

This section describes the experimental procedures used to acquire the
measurements supplied to the material constitutive inference framework. As already described, two
different modalities of experimental observations are utilized for this study. The first is the global
force--displacement ($F$--$D$) curve, where $F$ denotes the punch load
recorded by the load cell and $D$ denotes the punch displacement obtained
from the machine crosshead signal after correction for machine--fixture
compliance. The load and displacement signals were acquired synchronously
throughout the test. The second is the out-of-plane displacement field
$w(x,y)$ measured on the bottom surface of the clamped specimen by the
stereoscopic DIC system at a prescribed loading stage. The spatial
coordinates $(x,y)$ are defined in the undeformed specimen plane, with the
origin located at the punch axis. The selection of the loading stage at which
$w(x,y)$ is extracted, and its subsequent reduction to a compact feature
representation, are described in Section~\ref{sec:features}. The present
section describes only the physical acquisition of the $F$--$D$ and $w(x,y)$ datasets.

\subsection{Specimen preparation}
\label{sec:spt_protocol}

AA6111-T4 aluminum sheet alloy, used extensively in the automotive industry
\cite{kukielka2020aa6111,MORADI2026138,sarkar2004aa6111}, was selected for this study. SPT samples were cut from sheets by wire
electrical discharge machining (WEDM) at low discharge energy to
minimize heat-affected zone formation; following ASTM E3205-20
\cite{ASTM_E3205_20}, the nominal specimen dimensions were
10~mm~$\times$~10~mm~$\times$~0.5~mm. The specimen thickness was measured at
three locations per specimen, with variations within $\pm 0.005$~mm. Specimens were subsequently polished using 800- and 1200-grit sandpaper to
improve surface flatness and remove residual WEDM recast material. Independent tensile tests on the same sheet yielded a Young's modulus of
70~GPa and a 0.2\% offset yield strength of 157~MPa. These values are used
only as reference properties for comparison with the inferred posteriors; these were not exposed in any manner to the Bayesian inference protocol.

For DIC, a uniform matte white base coat (water-soluble acrylic paint,
cured for 1~hour under ambient conditions) was applied to the bottom
surface, followed by a high-contrast stochastic black speckle pattern
deposited with an Iwata Eclipse HP-CS airbrush at 2.0~bar and a
stand-off distance of 30~cm.  The resulting pattern exhibited a
characteristic speckle diameter of approximately 3--7~pixels and a
coverage fraction of approximately 50\%, within the range established
to provide adequate grayscale variation and robust subset correlation
\cite{seyedmahmoud2026sequential}.

\subsection{SPT and DIC acquisition}
\label{sec:dic}

\subsubsection{Mechanical setup and loading}

\begin{figure}[htbp]
\centering
\includegraphics[width=0.7\textwidth]{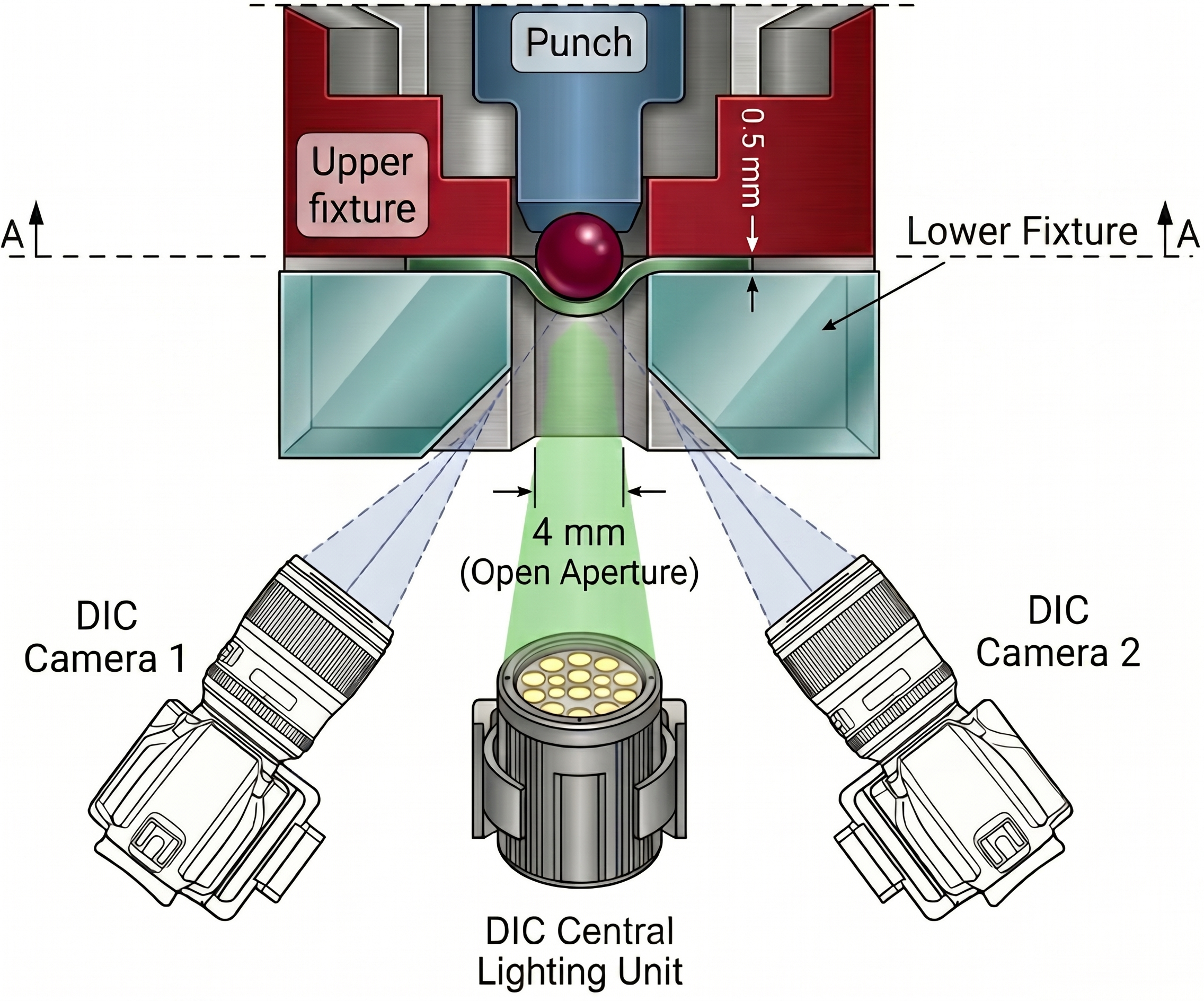}
\caption{Experimental setup: small punch test fixture with the
direct-view stereoscopic DIC system.  The punch and die assembly,
clamped specimen, two cameras with 100~mm lenses, and LED
illumination directed at the speckled bottom surface through the die
opening are shown.}
\label{fig:setup}
\end{figure}

Tests were performed on a servo-controlled electromechanical Zwick
testing machine equipped with a 2.5~kN hardness head.  The SPT fixture
followed the geometry of ASTM E3205-20 \cite{ASTM_E3205_20}: upper and
lower hardened tool-steel dies with a 4~mm circular opening, clamped by four symmetrically arranged bolts to provide a nominal clamping
force of approximately 8~kN. The punch consisted of a cylindrical
steel rod terminated with a 2.4~mm-diameter tungsten carbide ball,
aligned with the machine axis through a precision collet and
low-clearance bushing to suppress lateral motion.  The lower die
incorporated a conical cut-out to provide the DIC cameras with a
direct, unobstructed view of the bottom surface of the clamped specimen
(Figure~\ref{fig:setup}).  Silicone-based lubricant was applied to the
punch--specimen interface to reduce friction and promote repeatable
$F$--$D$ behaviour \cite{ASTM_E3205_20}.

All tests were conducted in displacement control at a constant punch rate of
0.5~mm/min, consistent with quasi-static SPT practice. Loading was continued until a load drop exceeding 20\% of the maximum load
$F_\mathrm{m}$ was observed, indicating specimen failure.

\subsubsection{Stereo DIC system}

Full-field three-dimensional surface deformation of the specimen bottom
surface was measured using a stereoscopic DIC system comprising two
Allied Vision Prosilica GT2050 cameras (2048~$\times$~1088~pixel
resolution) fitted with Kowa 100~mm fixed-focal-length lenses, arranged
symmetrically about the punch axis at a stereo angle of 29.4$^\circ$.
A direct-view configuration was adopted in which both cameras observed the speckled bottom surface through the die opening. This configuration avoids the additional alignment, optical-path, and distortion errors associated with intermediate mirror arrangements.  Synchronized image pairs were acquired at 1~frame/500 ms,
compatible with the imposed punch displacement rate.

Stereo calibration was performed prior to testing using a planar
dot-grid target ($9\times 9$ array, 0.45~mm dot spacing) viewed at
locations spanning the anticipated depth range of the specimen.
Calibration and correlation were performed in VIC-3D (Correlated
Solutions); the calibration reprojection error was 0.06~pixels and
the spatial sampling in the region of interest was approximately
372~pixels/mm.  Correlation employed a 31~$\times$~31~pixel subset
with a 3-pixel step and the normalized cross-correlation criterion.
With this spatial sampling, the 3-pixel step corresponds to a DIC
correlation-grid spacing of $3/372 = 0.0081$~mm, or approximately
8~$\mu$m, on the specimen surface.
The three-dimensional displacement components at each correlation node were
obtained from every image pair. In the Bayesian calibration framework, only the out-of-plane component $w(x,y)$ is
used from the DIC measurement.
A summary of the DIC system and correlation parameters is given in
Table~\ref{tab:dic_params}.

\begin{table}[htbp]
\centering
\caption{Summary of the 3D DIC system and correlation parameters.}
\label{tab:dic_params}
\begin{tabular}{ll}
\toprule
Quantity & Value \\
\midrule
Camera model        & Allied Vision Prosilica GT2050 \\
Sensor resolution   & 2048~$\times$~1088~pixels \\
Lens                & Kowa LM100JC1MS, 100~mm focal length \\
Spatial sampling    & $\approx$372~pixels/mm \\
Stereo angle        & 29.4$^\circ$ \\
Image acquisition   & 1~frame/500 ms \\
Correlation software & VIC-3D (Correlated Solutions, Inc.) \\
Subset size         & 31~$\times$~31~pixels \\
Subset step size    & 3~pixels \\
Correlation criterion & Normalized cross-correlation \\
Calibration reprojection error & 0.06~pixels \\
\bottomrule
\end{tabular}
\end{table}

\subsubsection{Displacement measurement accuracy}
\label{sec:dic_accuracy}

The accuracy of the out-of-plane displacement measurement was assessed with
a rigid-body motion test. The stereo camera assembly was translated relative
to the stationary specimen--fixture assembly by known displacements of 0.5,
1.0, and 1.5~mm along the viewing direction using a linear stage. At each
displacement, the mean DIC-measured out-of-plane displacement
$\langle w_\mathrm{DIC}\rangle$ was compared with the stage reference
$w_\mathrm{ref}$ (Table~\ref{tab:dic_validation}). The maximum absolute
error was 8.7~$\mu$m, corresponding to 0.58\% of the imposed displacement.
The system noise floor, measured from stationary images, was 0.06~$\mu$m.

\begin{table}[htbp]
\centering
\caption{Rigid-body validation of the out-of-plane DIC displacement
measurement. Known displacements $w_\mathrm{ref}$ were imposed along the
viewing direction and compared with the mean DIC-measured displacement
$\langle w_\mathrm{DIC}\rangle$. The signed error is defined as
$\Delta w=\langle w_\mathrm{DIC}\rangle-w_\mathrm{ref}$.}
\label{tab:dic_validation}
\begin{tabular}{cccc}
\toprule
$w_\mathrm{ref}$ ($\mu$m) & $\langle w_\mathrm{DIC}\rangle$ ($\mu$m) &
$\Delta w$ ($\mu$m) & $|\Delta w| / w_\mathrm{ref}\times 100$ (\%) \\
\midrule
500 & 499.9 & $-0.1$ & 0.02 \\
1000 & 998.5 & $-1.5$ & 0.15 \\
1500 & 1491.3 & $-8.7$ & 0.58 \\
\bottomrule
\end{tabular}
\end{table}

\section{Methods}
\label{sec:methods}

\begin{figure*}[htbp]
\centering
\includegraphics[width=\textwidth]{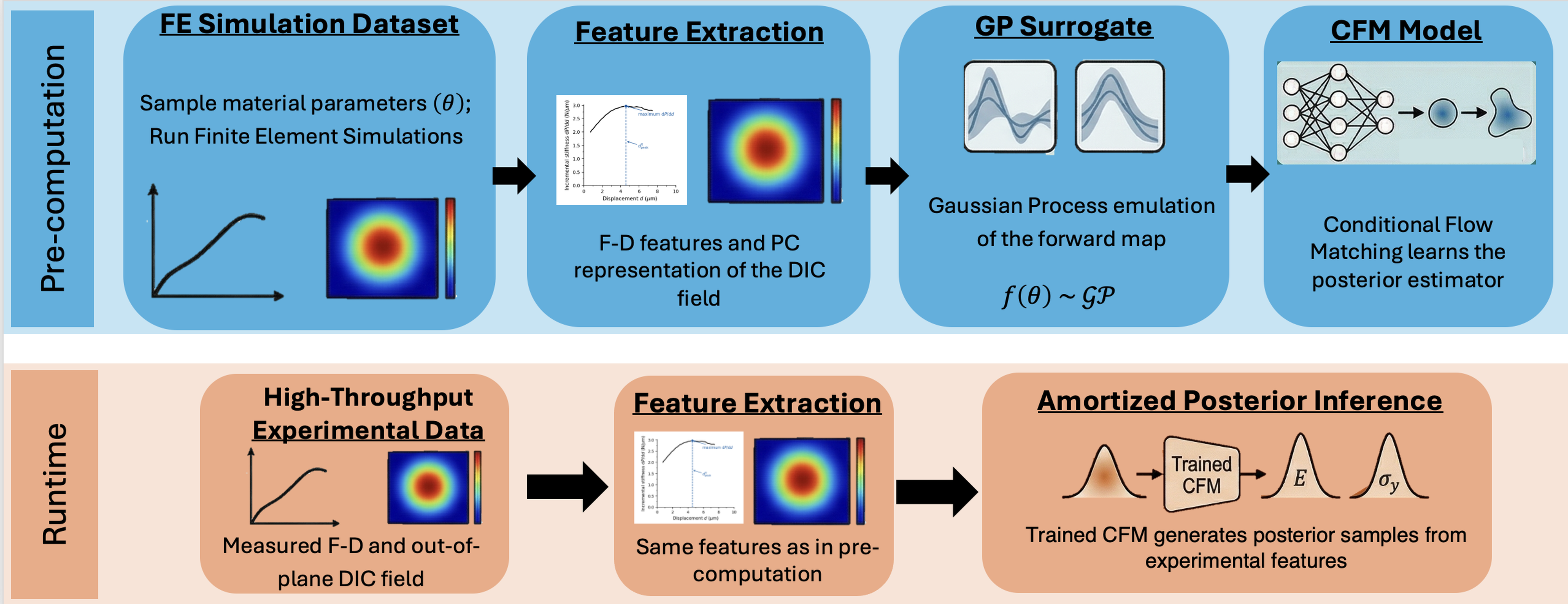}
\caption{Overview of the two-stage inference workflow.
\textit{Top (pre-computation):} finite element simulations are converted
into compact features that are used to train a surrogate model and a
posterior estimator. \textit{Bottom (runtime):} the experimental
$F$--$D$ and DIC measurements are processed using the same
feature representation and supplied to the trained CFM model to generate
posterior samples of the material parameters.}
\label{fig:workflow}
\end{figure*}

This section describes the two-stage amortized posterior inference workflow
used to infer constitutive parameters from the experimental SPT observations.
In the pre-computation stage, FE simulations are run over a Latin Hypercube
design of the inputs $(E, \sigma_y, t)$, their responses are reduced to
compact features, and these features are used to train a Gaussian process
(GP) surrogate and a Conditional Flow Matching (CFM) posterior estimator. At
runtime, the experimental $F$--$D$ curve and DIC displacement
field are reduced to the same features, which are passed, together with the
measured specimen thickness, to the trained CFM estimator to generate
posterior samples. Figure~\ref{fig:workflow} summarizes the pipeline. The
remainder of this section follows the same sequence.

\subsection{FE simulation dataset}
\label{sec:fe_dataset}

\begin{figure}[htbp]
\centering
\begin{subfigure}[b]{\columnwidth}
    \centering
    \includegraphics[width=0.75\columnwidth]{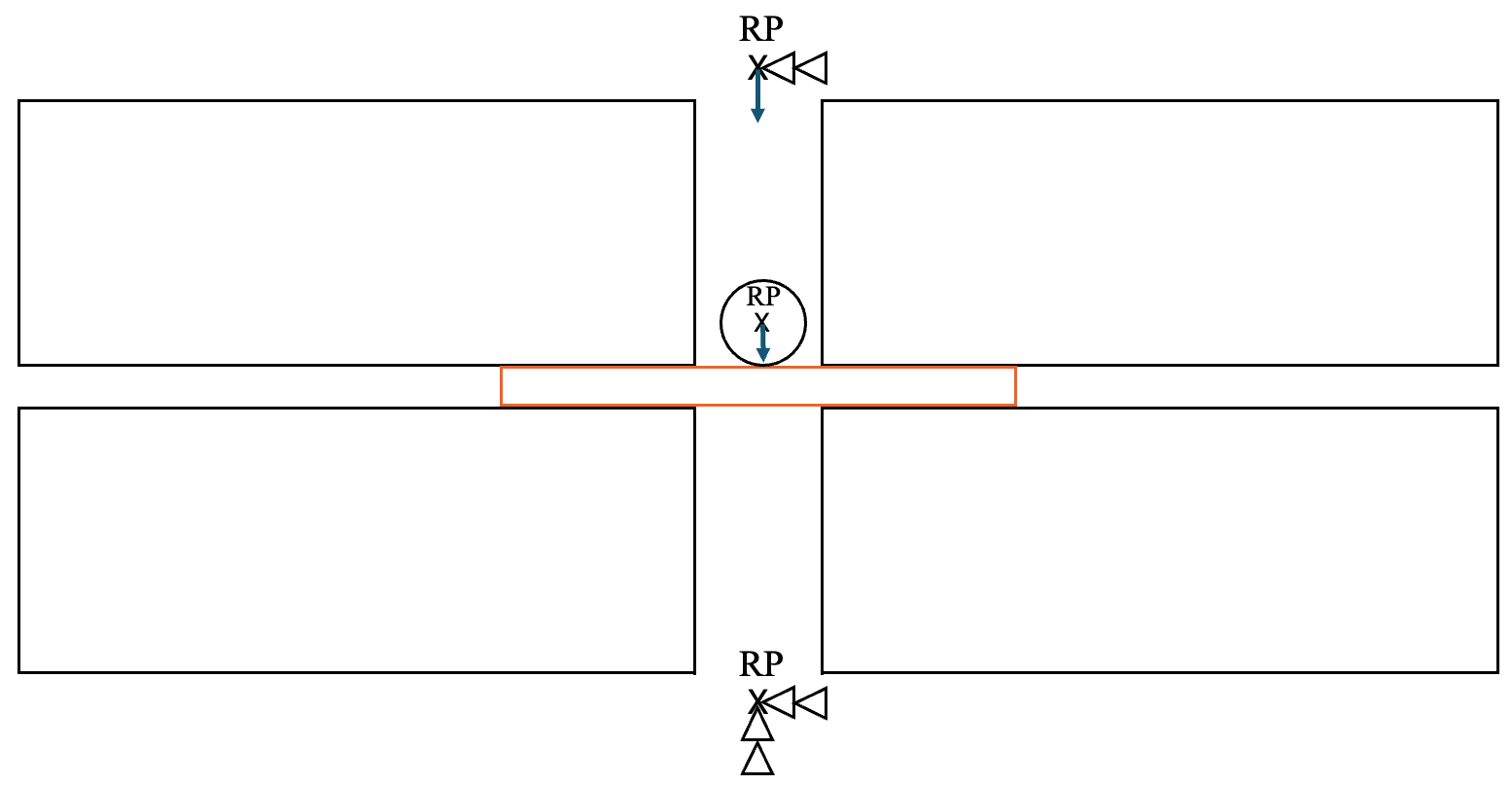}
    \caption{}
    \label{fig:mesh_bc}
\end{subfigure}
\vspace{0.5em}
\begin{subfigure}[b]{\columnwidth}
    \centering
    \includegraphics[width=\columnwidth]{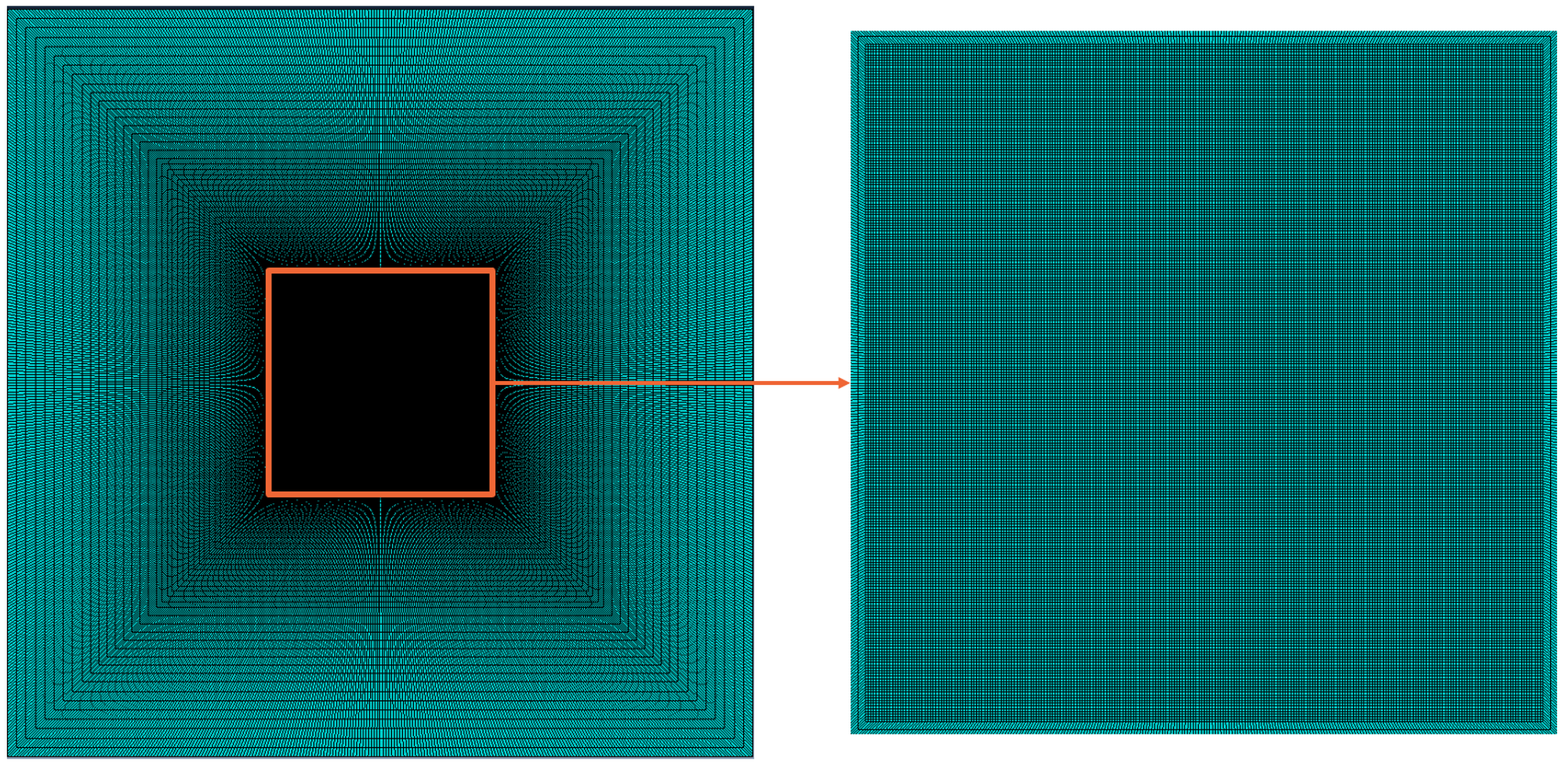}
    \caption{}
    \label{fig:mesh_mesh}
\end{subfigure}
\caption{Finite element model of the SPT specimen: (a) Boundary conditions. The punch reference point (RP) is
    loaded in displacement control (downward arrow). The upper die RP is
    held fixed after the clamping step. The lower die RP is fully fixed.
    The specimen (orange) sits between the dies. (b) FE mesh: full $10\times10$~mm domain (left) and zoom into
    the $3\times3$~mm DIC region of interest (right). The structured
    C3D8 mesh is locally refined toward the punch-contact region; within
    the region of interest the bottom-surface element faces form a
    regular rectangular grid.}
\label{fig:mesh}
\end{figure}

A finite element model of the SPT was implemented in
Abaqus/Standard~\cite{abaqus2022}. The specimen was modeled as a
$10\times10$~mm square sheet with a constant thickness $t$ (this geometric parameter
was varied as needed). A three-dimensional (3-D) model was developed and
used in this study. Although a 2-D
axisymmetric model is adequate to predict the global $F$--$D$
response, a 3-D model was used here to enable direct
comparison between the FE-predicted bottom-surface displacement field and
the DIC measurement over the exact same two-dimensional region of interest. 
Following an approach we successfully
employed in an earlier study~\cite{seyedmahmoud2026sequential}, the
specimen was discretized using a structured mesh of eight-node hexahedral
(C3D8) elements, with local refinement toward the punch-contact region
and the die-opening edge and coarser elements toward the specimen
boundaries (Figure~\ref{fig:mesh}). 
Within the $3\times3$~mm DIC region of interest, the bottom-surface
element faces form a regular rectangular grid with a node spacing of
approximately 0.008 mm, matching precisely the DIC correlation-grid spacing 
defined in Section~\ref{sec:dic}. After coordinate alignment, the FE-predicted and DIC-measured
displacement fields are compared by direct node-to-node correspondence over
the common ROI, without interpolation. A mesh convergence
study confirmed that the $F$--$D$ response over the early response
interval and the peak out-of-plane displacement
over the DIC region of interest were insensitive to further refinement;
after halving the element size in the contact region, the relative change
in both quantities was less than 0.6\%.

The hemispherical punch and the upper and lower dies were modeled as
analytical rigid bodies, each controlled by an associated reference point
(Figure~\ref{fig:mesh_bc}). The contact geometry reproduced the
ASTM~E3205-20 tooling dimensions, with punch radius $r_p = 1.2$~mm and
die opening radius $r_d = 2$~mm. The
lower-die reference point was fixed. The upper die was displaced through 
its reference point in a preliminary
clamping step until the reaction force reached approximately 8~kN,
consistent with the experimental fixture preload described in
Section~\ref{sec:spt_protocol}. This upper-die displacement was then held
fixed during punch loading. The punch was loaded in displacement control
through its reference point to a maximum displacement of $D = 0.1$~mm. 

The deformable specimen was modeled using an elastic--perfectly plastic
(EPP) constitutive law, so that the material property calibration in this study
focused exclusively on the elastic stiffness and initial yield strength. 
The elastic response is defined by
\begin{equation}
    \bm{\sigma} = \mathbb{C}(E,\nu):\bm{\varepsilon}^{e},
    \label{eq:hooke}
\end{equation}
where $E$ is Young's modulus, $\nu$ is Poisson's ratio, and
$\mathbb{C}$ is the isotropic elastic stiffness tensor. Plastic yielding is
described by the von Mises yield function
\begin{equation}
    f(\bm{\sigma},\sigma_y)
    =
    \sigma_\mathrm{eq} - \sigma_y \leq 0,
    \label{eq:yield}
\end{equation}
where $\sigma_\mathrm{eq}$ is the von Mises equivalent stress and
$\sigma_y$ is the initial yield strength. Associated plastic flow rule is used:
\begin{equation}
    \dot{\bm{\varepsilon}}^{p}
    =
    \dot{\lambda}
    \frac{\partial f}{\partial \bm{\sigma}} .
    \label{eq:flow_rule}
\end{equation}
No hardening parameters are included. This choice deliberately limits the
present study to a controlled two-parameter calibration problem for which
independent tensile reference values of $E$ and $\sigma_y$ are available:
\begin{equation}
    \bm{\theta}
    =
    \begin{bmatrix}
    E & \sigma_y
    \end{bmatrix}^{\top}.
    \label{eq:theta_def}
\end{equation}

Table~\ref{tab:fe_inputs} summarizes the values and ranges used for the FE
model inputs. The training dataset was generated using Latin Hypercube
Sampling (LHS) of $N=300$ parameter sets~\cite{mckay1979lhs}. Young's modulus
$E$ and yield strength $\sigma_y$ (the constitutive parameters to be estimated) were
sampled over ranges broader than the nominal properties of AA6111-T4 so that
the surrogate and posterior estimator were not trained only near the
reference material. Specimen thickness $t$ was included as a measured
geometric input because it varied among specimens due to the preparation
procedure described in Section~\ref{sec:spt_protocol}. Its sampled range is
consistent with the ASTM~E3205-20 nominal thickness
specification~\cite{ASTM_E3205_20}. Poisson's ratio was held constant at a
value representative of aluminum alloys~\cite{ashby2011materials}, and the
punch--specimen friction coefficient was held constant at a value consistent
with lubricated SPT contact conditions reported in FE
studies~\cite{leclerc2021spt}. FE simulation checks confirmed that varying $\nu$ over
$0.27$--$0.33$ and $\mu$ over $0.01$--$0.10$ changed the FE outputs of interest to this study by less than 1\%.

\begin{table}[htbp]
\centering
\caption{FE model inputs used in the simulation dataset.}
\label{tab:fe_inputs}
\begin{tabular}{lll}
\toprule
Quantity & Treatment & Value or range \\
\midrule
Young's modulus, $E$ & inferred & $60$--$200$~GPa \\
Yield strength, $\sigma_y$ & inferred & $120$--$500$~MPa \\
Specimen thickness, $t$ & measured/geometric input & $0.45$--$0.55$~mm \\
Poisson's ratio, $\nu$ & constant & $0.30$ \\
Friction coefficient, $\mu$ & constant & $0.10$ \\
\bottomrule
\end{tabular}
\end{table}

As already noted, our interest here is exclusively on the early portion of the $F$--$D$ curves that are sensitive only to the values of $E$ and $\sigma_y$. Consequently, we need to suitably truncate the $F$--$D$ curve to just before the coalescence of a through-thickness plastic region (see the middle plot in Figure~\ref{fig:intro_coupling_peeq}). FE simulations have indicated that the displacement value where this coalescence occurs varies significantly with the values of $E$ and $\sigma_y$. As a result, the $F$--$D$ curves to be analyzed for estimating the values of $E$ and $\sigma_y$ will correspond to different displacement truncation values. Our attempts using a fixed-displacement
cutoff produced significantly degraded estimates of the materials properties of interest. 

\begin{figure}[htbp]
\centering
\includegraphics[width=\columnwidth]{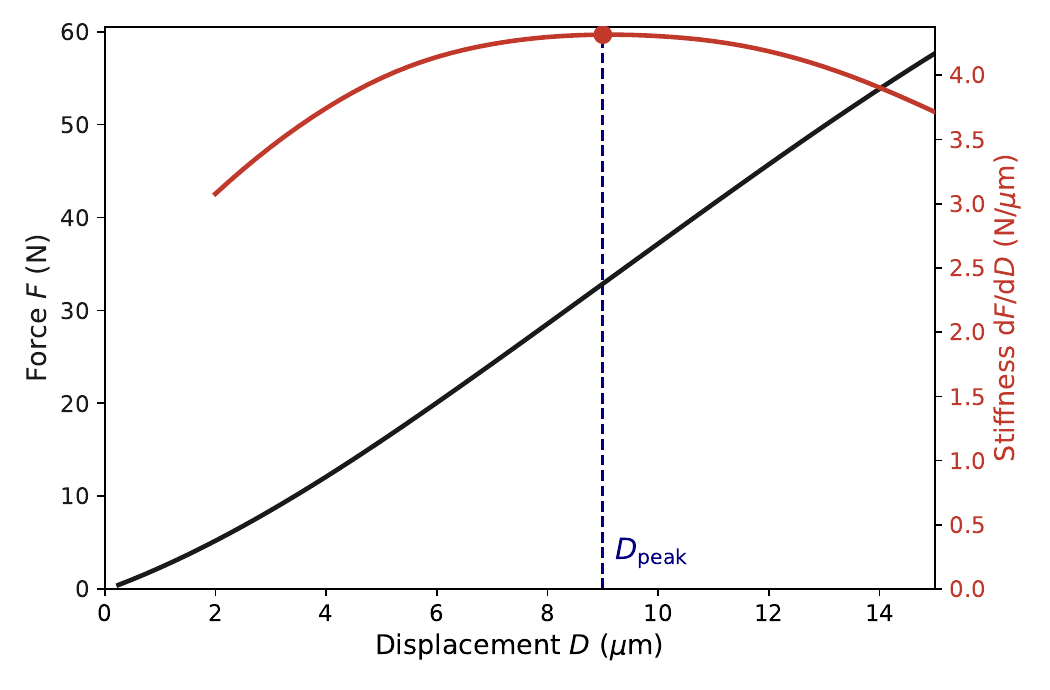}
\caption{Truncation of the $F$--$D$ curve for the estimation of $E$ and $\sigma_y$.  The force $F$ (black, left axis) and the stiffness $\mathrm{d}F/\mathrm{d}D$ (red, right axis) are
plotted against punch displacement. The displacement
$D_\mathrm{peak}$ corresponding to the maximum in the $\mathrm{d}F/\mathrm{d}D$
was used to truncate the $F$--$D$ curves for the analyses in this study.}
\label{fig:dpeak_def}
\end{figure}

Previous studies have used gradient- and curvature-based
characteristics of the initial SPT force--displacement response to identify the elastic-plastic transition ~\cite{calafchica2018slope,hahner2019determining}. Building on these prior efforts, we identify the displacement truncation value in our study using the
stiffness $\mathrm{d}F/\mathrm{d}D$ response. Specifically, we observed that the stiffness--displacement response exhibited a maximum close to the through-thickness plastic coalescence
(see Figure~\ref{fig:dpeak_def}). The displacement corresponding to the maximum stiffness is denoted as $D_\mathrm{peak}$, and has been used as a truncation displacement value for the $F$--$D$ curves analyzed in this study. 

Locating $D_\mathrm{peak}$  reliably from both the experimental and FE simulated $F$--$D$ curves required the development of a stable numerical protocol. For this purpose, we excluded the very early portion of the $F$--$D$ curve (i.e., below $D=0.5~\mu\mathrm{m}$) as this portion was found to be noisy in both experimental and simulated curves. The
remaining $F$--$D$ curve was smoothed using a Savitzky--Golay
filter~\cite{savitzky1964smoothing} with a third-order polynomial and an
11-point window before numerical differentiation.

\subsection{Feature extraction}
\label{sec:features}

Two specific responses from both the experiments and the FE simulations are analyzed in this study. 
The first is the truncated $F$--$D$ curve, sampled at
approximately 400 displacement points, and the second is the bottom-surface
out-of-plane displacement field at $D_\mathrm{peak}$ over the DIC region of interest, represented by
approximately 140,000 nodal values. These sampled representations are too
high-dimensional to model directly with a GP surrogate trained on 300
simulations. Compact, low-rank features of the simulated and experimental
responses were therefore constructed to support the surrogate modeling and
posterior inference that follow.

\subsubsection{$F$--$D$ features}
\label{sec:fd_features}

The truncated $F$--$D$ curves were observed to fit well to a power-law approximation:
\begin{equation}
    F = A D^n .
    \label{eq:powerlaw_fd}
\end{equation}
Further, it was observed that the value of the exponent $n$
varied only narrowly from 1.14 to 1.16 across the entire simulation dataset.
Therefore, it was kept fixed at $n = 1.15$; only the value of $A$ was estimated
for each truncated $F$--$D$ curve. Each truncated SPT $F$--$D$
curve was represented using two features: the fitted
coefficient $A$ and the truncation displacement $D_\mathrm{peak}$. The mean $R^2$  value across all 300 simulations used for the training dataset was 0.99, justifying the selection of the power-law fit for the present study.

\subsubsection{DIC Features}
\label{sec:dic_features}

As already described earlier, the DIC-measured bottom-surface out-of-plane displacement field
$w(x,y)$ at $D_\mathrm{peak}$ is utilized as an additional source of information for estimating the materials properties of interest. Unlike the $F$--$D$ curves that reflect an overall sample-averaged elastic-plastic response, the DIC measurements contain spatially resolved information from the different regions of the sample undergoing different levels of elastic and plastic deformation.  Consequently, it is expected to be very informative in the estimation of both $E$ and $\sigma_y$. The
displacement magnitude at $D_\mathrm{peak}$ is also well above the validated
DIC noise floor, allowing the field to be measured reliably.

In this work, the DIC-measured and FE-simulated $w(x,y)$ at $D_\mathrm{peak}$ were both captured into a flattened vector of size 139,129. As already described, we ensured an exact  correspondence between the DIC measurements and the FE mesh to accomplish this task. Low-rank representations of the displacement field were established by performing a principal component analysis (PCA)~\cite{wold1987pca} on the 300 simulated displacement fields obtained in the training set.  The first three PC scores, denoted $\mathrm{PC}_1$--$\mathrm{PC}_3$, were
retained as the DIC features for this study as they accounted
for more than 99\% of the variance in the training set.
The PCA basis obtained from the FE simulation training dataset is also applied on the DIC-measured displacement fields.

Combining the two $F$--$D$ features and the three DIC PC scores, we arrive at the feature vector used for surrogate training and posterior inference in this study:
\begin{equation}
    \mathbf{y}
    =
    \bigl(
    \underbrace{A,\;D_\mathrm{peak} }_{F\text{--}D},\;
    \underbrace{\mathrm{PC}_1,\;\mathrm{PC}_2,\;\mathrm{PC}_3}_{\mathrm{DIC}}
    \bigr)^\top
    \in \mathbb{R}^{5}.
    \label{eq:feature_vec}
\end{equation}

\subsection{GP surrogate}
\label{sec:gp_surrogate}

A Gaussian process (GP) surrogate~\cite{rasmussen2006gp} is constructed to
emulate the mapping from the constitutive parameters $\bm{\theta}$ defined in Eq.~\eqref{eq:theta_def}
and specimen thickness $t$ to the features $\mathbf{y}$ identified in Eq.~\eqref{eq:feature_vec}.
This mapping is aimed at replacing the FE model in the generation of the large
number of synthetic parameter--feature pairs required to train the posterior
estimator and in the posterior predictive evaluations.  The GP models for this study were implemented in
GPyTorch~\cite{gardner2018gpytorch}. Five independent single-output GPs were
constructed, one for each feature in $\mathbf{y}$. Each GP predicts one component of $\mathbf{y}$, denoted $y$, using a
linear mean function,
\begin{equation}
    \mathbb{E}\!\left[y(\mathbf{x})\right] = \mathbf{w}^{\top}\mathbf{x} + b,
    \label{eq:gp_mean}
\end{equation}
and its covariance is a scaled radial basis function kernel with automatic
relevance determination (ARD),
\begin{equation}
    \operatorname{Cov}\!\left[y(\mathbf{x}),y(\mathbf{x}')\right]
    =
    \sigma^2
    \exp\!\left[
    -\frac{1}{2}
    \sum_{i=1}^{3}
    \frac{(x_i-x_i')^2}{\ell_i^2}
    \right],
    \label{eq:rbf_kernel}
\end{equation}
where $\mathbf{x}=(E,\sigma_y,t)^\top$ are the GP inputs, $\ell_i$ is the
length-scale associated with input $x_i$, and $\sigma^2$ is the kernel
variance. A short length-scale $\ell_i$ indicates that $y$ varies rapidly with
input $x_i$;  the ARD parameterization allows each GP to assign a different
effective sensitivity to $E$, $\sigma_y$, and $t$. Each GP therefore has eight
hyperparameters, namely the mean weights $\mathbf{w}$ and bias $b$, the three
length-scales, and the kernel variance $\sigma^2$,
which are estimated by maximizing the marginal log-likelihood. All inputs and
feature outputs are Z-score normalized using training-set statistics before
fitting. Because the FE outputs are deterministic, the GP observation-noise
variance is not fitted; a small fixed value of $10^{-4}$ 
is added to the kernel-matrix diagonal solely for numerical stability, so the
resulting predictive variance represents surrogate interpolation uncertainty
rather than experimental measurement noise.

This parameterization is deliberately parsimonious. Across the five GPs it
involves only 40 hyperparameters, in contrast to the several hundred that
would be required to emulate the full $F$--$D$ curves and
bottom-surface displacement fields directly, and it achieves this without
loss of predictive accuracy (Section~\ref{sec:surrogate_fidelity}).

\subsection{CFM model}
\label{sec:cfm_inference}

After feature extraction, inference is posed as estimation of the posterior
distribution of the constitutive parameters conditioned on the measured
features and specimen thickness,
\begin{equation}
    p(\bm{\theta} \mid \mathbf{y},\,t)
    \;\propto\;
    p(\mathbf{y} \mid \bm{\theta},\,t)\,p(\bm{\theta}),
    \label{eq:bayes}
\end{equation}
where $\mathbf{y}\in\mathbb{R}^{5}$ is the feature vector defined in
Eq.~\eqref{eq:feature_vec}, $t$ is the measured specimen thickness, and
$p(\bm{\theta})$ is the uniform prior over
\begin{equation}
    E \in [60,\,200]\ \mathrm{GPa},
    \qquad
    \sigma_y \in [120,\,500]\ \mathrm{MPa}.
    \label{eq:prior_bounds}
\end{equation}
These bounds coincide with the $E$ and $\sigma_y$ ranges listed in
Table~\ref{tab:fe_inputs}. Equation~\eqref{eq:bayes} states the
formal Bayesian target. The likelihood term is not evaluated directly;
instead, the conditional posterior is estimated from synthetic
parameter--feature pairs.

The posterior $p(\bm{\theta}\mid\mathbf{y},t)$ is approximated using
Conditional Flow Matching (CFM)~\cite{lipman2023flow}, a simulation-based
neural posterior estimator within the broader class of likelihood-free
inference methods~\cite{cranmer2020frontier,papamakarios2021normalizing}.
CFM learns a continuous transport from a tractable reference distribution to
the target posterior conditioned on the feature vector and specimen
thickness. In this setting, the dependence among the conditioning features
is learned from the simulated joint distribution of parameters and features,
rather than imposed through an analytical likelihood; the conditioning input
can therefore combine quantities from different measurement modalities
without an explicit cross-modality covariance model. Compared with
adversarial generative models, CFM avoids min--max training; compared with
discrete normalizing flows, it does not require explicitly invertible
network architectures~\cite{albergo2023stochastic}. In practical terms, the
trained model takes the measured feature vector as input and returns samples
of the constitutive parameters consistent with that measurement. Although
normalizing flows and related transport models have been used to accelerate
Bayesian inverse problems in several scientific
domains~\cite{sherki2025cfm,dasgupta2026cfm}, their application to
 constitutive calibration using multimodal data (here SPT--DIC) has not yet been demonstrated.

The CFM training set combines two sources of parameter--feature pairs. The
base set consists of the 210 FE training simulations, each contributing its
GP input $\mathbf{x}=(\bm{\theta}, t)$ and extracted feature vector $\mathbf{y}$. This
base set is augmented during training with synthetic pairs constructed by
sampling $\bm{\theta}$ from the prior ranges and $t$ from the thickness
range listed in Table~\ref{tab:fe_inputs} and evaluating the GP surrogate,
as described below. Measurement uncertainty in the extracted features is
represented by independent Gaussian noise terms in normalized feature
coordinates,
\begin{equation}
    p(\mathbf{y}\mid\bm{\theta},t)
    =
    \prod_{k\in\mathcal{K}}
    \mathcal{N}
    \bigl(
    y_k;\;
    \hat{y}_k(\mathbf{x}),\;
    \nu_k
    \bigr),
    \label{eq:obs_model}
\end{equation}
where $\mathcal{K}$ is the set of conditioning features used in a given CFM
model, $\hat{y}_k(\mathbf{x})$ is the GP predictive mean for
feature $k$, and $\nu_k$ is the corresponding feature-level noise variance.
The variances $\nu_k$ are treated as fixed constants in the present study.
For the DIC-derived principal-component scores, these variances are obtained
by propagating the out-of-plane displacement noise floor established in
Section~\ref{sec:dic_accuracy} ($0.06~\mu\mathrm{m}$) through the PCA
projection. For the $F$--$D$
features, the variances are estimated by propagating load-cell and crosshead
displacement uncertainties through the feature-extraction procedure. The
noise levels are specified after normalizing each feature to zero mean and
unit standard deviation. When the propagated sensor uncertainty was very
small, the feature-noise standard deviation was set to a conservative
minimum floor of 0.01 in normalized units to prevent the CFM from being
trained on observations with implausibly low uncertainty.

The diagonal form in Eq.~\eqref{eq:obs_model} is used only to inject
feature-level uncertainty when generating synthetic training observations.
It does not impose independence on the inferred parameters, since posterior
correlations arise from the joint dependence of all features on the shared
constitutive parameters. Noise correlations between $F$--$D$ and DIC features are not
explicitly modeled in this generator, which is one motivation for assessing
posterior calibration through the diagnostics in
Section~\ref{sec:validation_strategy}. In this fixed-noise treatment, the
feature-level measurement variances are specified in advance rather than
inferred hierarchically. Placing a prior over the noise variances and
marginalizing them would produce a broader effective uncertainty model, but
is left for future work.

Equation~\eqref{eq:obs_model} is therefore used as a stochastic
training-data generator, not as a calibrated residual model for the
experimental system. After training, the CFM model directly maps
the conditioning vector $(\mathbf{y},t)$ to samples from the estimated
conditional posterior, so posterior sampling for a new specimen requires no
likelihood calls or MCMC chain.

Let
\begin{equation}
    \mathbf{z}
    =
    \frac{\bm{\theta}-\bm{\mu}_{\theta}}{\bm{\sigma}_{\theta}}
    \label{eq:zscore_theta}
\end{equation}
denote the normalized parameter vector, where $\bm{\mu}_{\theta}$ and
$\bm{\sigma}_{\theta}$ are the mean and standard deviation of the parameter
samples in the training design. The reference distribution is taken as a
standard normal distribution,
$\mathbf{z}_0\sim\mathcal{N}(\mathbf{0},\mathbf{I})$. For a target parameter
sample $\mathbf{z}_1$, CFM constructs a linear interpolation at random time
$\tau\sim\mathcal{U}(0,1)$,
\begin{equation}
    \mathbf{z}_{\tau}
    =
    (1-\tau)\mathbf{z}_0+\tau\mathbf{z}_1,
    \qquad
    \mathbf{v}^{*}
    =
    \mathbf{z}_1-\mathbf{z}_0 .
    \label{eq:cfm_interp}
\end{equation}
The neural velocity field
$\mathbf{v}_{\phi}(\mathbf{z}_{\tau},\tau,\mathbf{c})$, referred to here as
VelocityNet, is trained to regress the target velocity,
\begin{equation}
    \mathcal{L}(\phi)
    =
    \mathbb{E}_{\tau,\mathbf{z}_0,(\mathbf{z}_1,\mathbf{c})}
    \left[
    \left\|
    \mathbf{v}_{\phi}(\mathbf{z}_{\tau},\tau,\mathbf{c})
    -
    \mathbf{v}^{*}
    \right\|^2
    \right],
    \label{eq:cfm_loss}
\end{equation}
where $\mathbf{c}$ is the normalized conditioning vector containing the
available features and the measured specimen thickness.

Two CFM models are trained, differing only in their conditioning inputs. One
is conditioned on the $F$--$D$ features $A$ and $D_\mathrm{peak}$
with thickness $t$; the other additionally includes the three DIC
principal-component scores. Comparing the two isolates the contribution of the
DIC modality to parameter identifiability. Training separate models avoids
assigning placeholder values to unobserved DIC features and ensures that each
posterior is conditioned only on genuinely available measurements.

The VelocityNet architecture and training hyperparameters were selected by
monitoring the CFM training loss and posterior coverage on a validation subset
held out from the 210 training simulations, balancing expressivity against
overfitting on the finite simulation dataset. The 90 test-set simulations were
not used for this selection and therefore remain fully out-of-sample for the
calibration diagnostics. The selected values are summarized in
\ref{app:cfm_details}.

To include surrogate uncertainty in the CFM training distribution, GP
augmentation is performed during training. At each epoch,
$N_\mathrm{aug}=1000$ synthetic parameter--feature pairs are generated by
sampling parameter and thickness values uniformly within the sampling
ranges and are combined with the 210 base FE pairs. This value was selected to provide dense coverage of the
parameter space relative to the batch size of 128; preliminary checks
confirmed that increasing $N_\mathrm{aug}$ beyond this value produced no
measurable change in the trained posterior. This augmentation distribution combines the synthetic
feature-noise model in Eq.~\eqref{eq:obs_model} with the GP predictive
variance, so that each synthetic training pair reflects both feature-level
measurement uncertainty and surrogate interpolation uncertainty. For feature
$k$, the augmented feature value is sampled as
\begin{equation}
    y_{\mathrm{aug},k}
    \sim
    \mathcal{N}
    \left(
    \hat{y}_k(\mathbf{x}),
    \sigma^2_{\mathrm{GP},k}(\mathbf{x}) + \nu_k
    \right),
    \qquad k=1,\ldots,5,
    \label{eq:gp_aug}
\end{equation}
where $\sigma^2_{\mathrm{GP},k}$ is the latent GP predictive variance
associated with surrogate interpolation uncertainty, and $\nu_k$ is the
feature-level measurement variance defined above.
Sampling from this distribution, rather than from the GP mean alone,
propagates both surrogate interpolation uncertainty and feature-level
measurement uncertainty into the estimated posterior. This is intended to limit
overconfidence in regions that are sparsely covered by the FE training
design~\cite{seyedmahmoud2026sequential}.

At inference, independent samples
$\mathbf{z}_0^{(s)}\sim\mathcal{N}(\mathbf{0},\mathbf{I})$ are transported
to the posterior by integrating the learned conditional ODE,
\begin{equation}
    \frac{\mathrm{d}\mathbf{z}}{\mathrm{d}\tau}
    =
    \mathbf{v}_{\phi}(\mathbf{z},\tau,\mathbf{c}_\mathrm{obs}),
    \qquad
    \tau\in[0,1],
    \label{eq:cfm_ode}
\end{equation}
where $\mathbf{c}_\mathrm{obs}$ is the conditioning vector extracted from the
experimental feature vector and measured specimen thickness. The ODE is
integrated using a fourth-order Runge--Kutta scheme with 20 fixed steps, and
a total of $N_\mathrm{post}=20\,000$ posterior samples are generated for
each experiment using one batched integration. Once trained, the CFM
weights, GP surrogates, and PC basis remain fixed; for a new specimen, only
the conditioning vector changes. The full sampling algorithm and the
architecture and training hyperparameters are provided in
\ref{app:cfm_details}.

\subsection{Posterior validation strategy}
\label{sec:validation_strategy}

We organize posterior validation around two questions, namely whether the DIC
modality improves parameter identifiability and whether the amortized
posterior estimator is statistically calibrated on held-out simulations.

The 300 FE simulations are partitioned into training and held-out test sets
using a 70/30 split. The surrogate's predictive accuracy is evaluated on the
90 held-out simulations and reported in Section~\ref{sec:surrogate_fidelity}.
For the calibration diagnostics described below, the GP surrogate and the CFM
estimator are trained only on the 210 training simulations, so that the 90
held-out cases remain fully out-of-sample; for the experimental inference, the
surrogate is retrained on all 300 simulations. These test-set assessments
therefore evaluate the surrogate form and, separately, CFM calibration under
the surrogate-based observation generator; they do not directly test the
models retrained for the experimental data.

The first set of diagnostics evaluates posterior contraction and predictive
consistency. The $F$--$D$-only and multimodal posteriors are
compared using the joint posterior shape, the widths of the 95\% highest
posterior density (HPD) credible intervals. Narrower marginal credible
intervals relative to the $F$--$D$-only case are used to assess
whether the additional modality improves identifiability.
Posterior predictive checks (PPCs) complement this comparison by testing
internal consistency. Posterior samples are propagated through the
stochastic observation model---the GP predictive distribution combined with
the feature-level noise---to obtain replicated feature vectors, which are
compared with the experimentally extracted feature vector. Agreement between the predicted and
measured features indicates that the inferred parameter samples are
consistent with the observations used for conditioning, whereas systematic
offsets indicate residual model--data discrepancy or feature-level
mismatch~\cite{gelman1996ppc}. The computation and interpretation of PPCs
are described in \ref{app:ppc}.

Posterior contraction alone is insufficient, because an overconfident
posterior may produce narrow credible intervals that fail to contain the true
parameter values. Calibration is therefore assessed using simulation-based calibration
(SBC)~\cite{talts2018sbc}, which checks whether the reported credible
intervals achieve their nominal frequentist coverage on held-out synthetic
cases with known parameter values. Because the synthetic observations are
generated from the same GP-based observation model used during CFM training,
SBC assesses the calibration of the CFM estimator under that surrogate-based
generator. It cannot detect GP bias relative to the FE outputs; surrogate
error is evaluated separately on the test set in
Section~\ref{sec:surrogate_fidelity}. The diagnostic also does not, by itself,
validate the fidelity of the FE model to the experimental system. The
computation and interpretation
of the rank histograms and empirical coverage curves used for
this assessment are described in \ref{app:sbc}. Each held-out case uses
2000 CFM posterior samples, which is sufficient for
rank-based diagnostics~\cite{talts2018sbc}, whereas 20,000 samples are
used for experimental inference to obtain stable marginal density estimates
and credible intervals. For these held-out cases, the true parameter values
are known from the Latin Hypercube design.

\section{Results and Discussion}
\label{sec:results}

We organize the results around the evidence needed to interpret the inferred
posteriors. We first evaluate surrogate accuracy on the FE test set
to determine whether the five extracted features can be emulated with
sufficient accuracy. The trained posterior estimator is then applied to a
representative AA6111-T4 SPT--DIC experiment, where we compare inference from
the $F$--$D$ features alone with inference from the combined
$F$--$D$ and DIC feature vector. Calibration diagnostics using
the test set assess whether the reported credible intervals have the
expected frequentist coverage. Posterior predictive
checks then examine whether the inferred parameters remain consistent 
with the measured responses. The tested
specimen has measured thickness $t=0.5$~mm.
Independent tensile measurements, $E=70$~GPa and $\sigma_y=157$~MPa, are used
only as external reference values.

\subsection{Surrogate accuracy}
\label{sec:surrogate_fidelity}

The CFM estimator is trained on feature vectors generated by the GP
surrogate. We first assess how accurately the surrogate represents the FE
feature map using the 30\% test set
($N_\mathrm{test}=90$), which was excluded from GP fitting.

This evaluation uses the five-dimensional feature vector defined in
Eq.~\eqref{eq:feature_vec}: the $F$--$D$ features $A$ and
$D_\mathrm{peak}$ and the three DIC PC scores evaluated at
$D_\mathrm{peak}$. As discussed in Section~\ref{sec:features}, these three
PC scores account for more than 99\% of the displacement-field variance in
the FE simulation dataset.

\begin{figure*}[htbp]
\centering
\begin{subfigure}[b]{0.45\textwidth}
    \includegraphics[width=\textwidth]{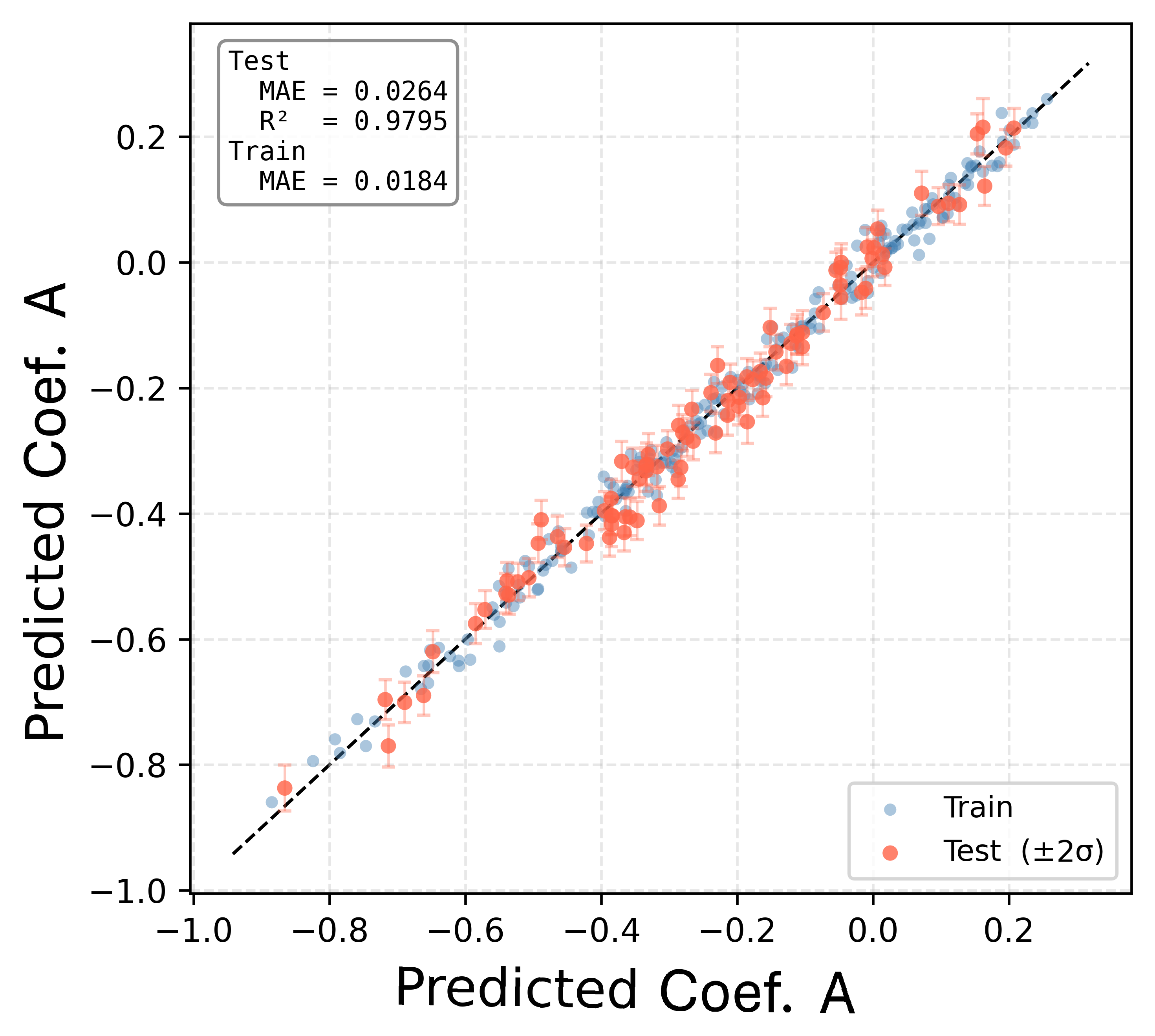}
    \caption{}
    \label{fig:parity_logA}
\end{subfigure}
\hfill
\begin{subfigure}[b]{0.45\textwidth}
    \includegraphics[width=\textwidth]{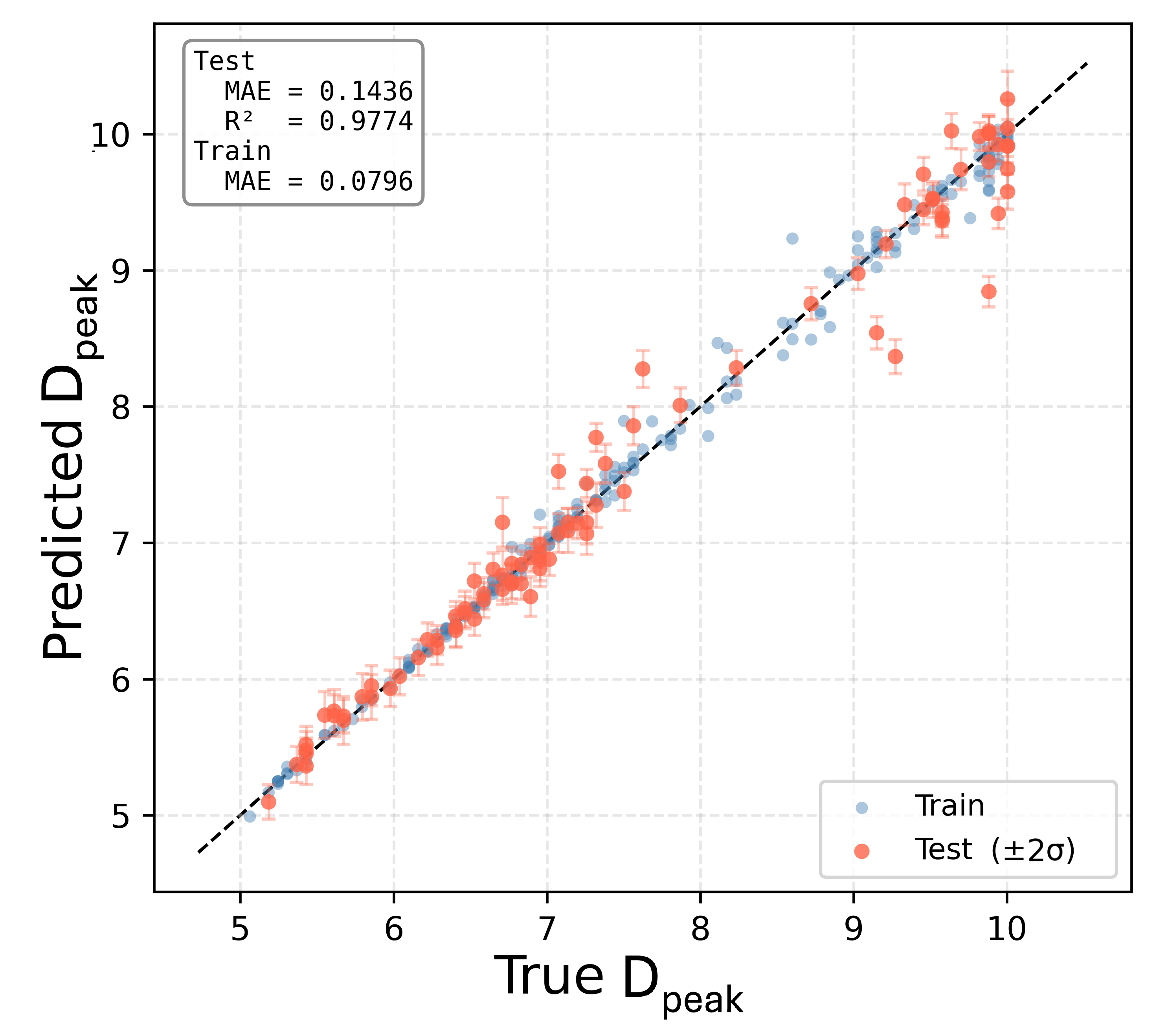}
    \caption{}
    \label{fig:parity_dpeak}
\end{subfigure}
\hfill
\begin{subfigure}[b]{0.45\textwidth}
    \includegraphics[width=\textwidth]{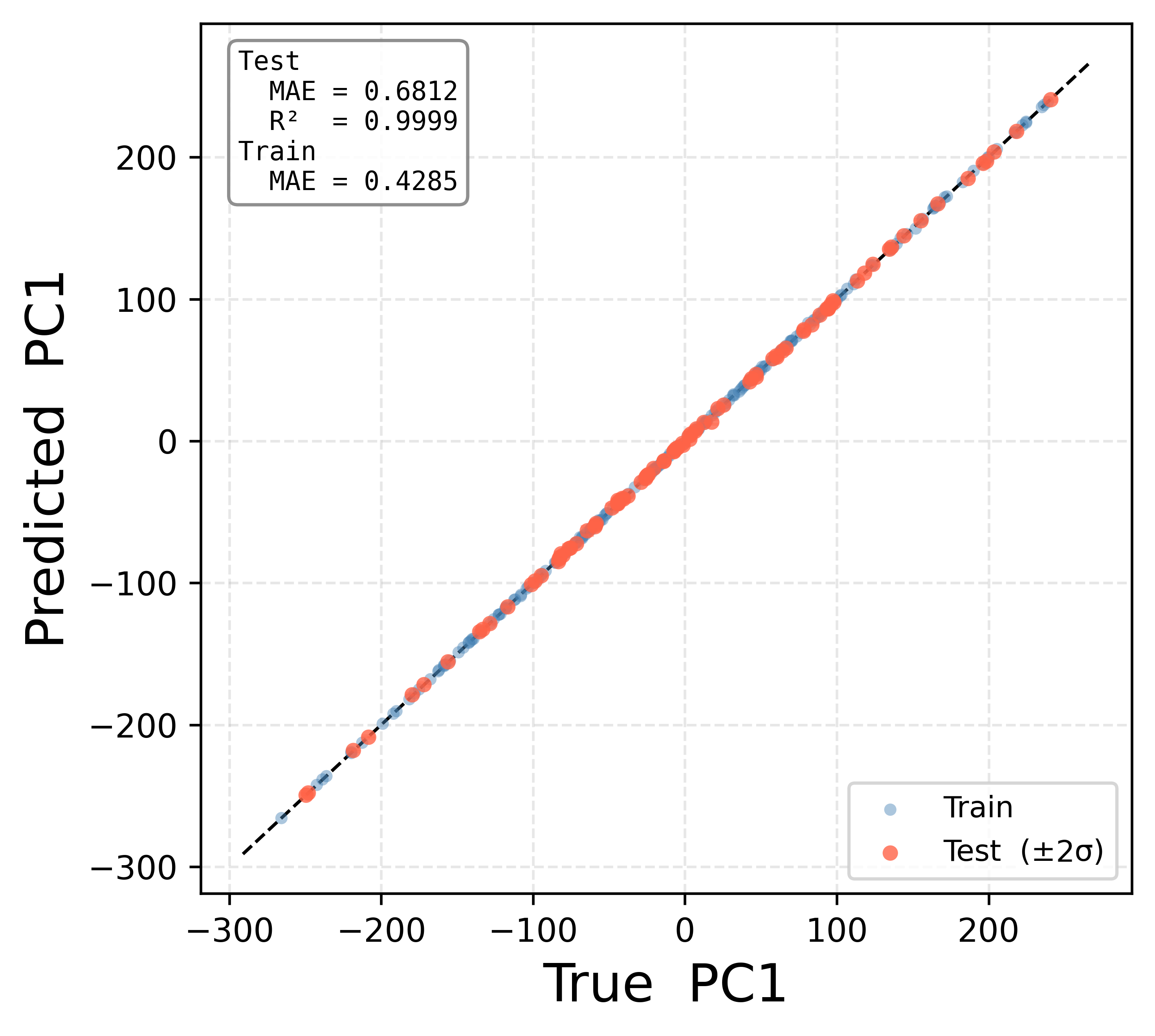}
    \caption{}
    \label{fig:parity_pc1}
\end{subfigure}
\hfill
\begin{subfigure}[b]{0.45\textwidth}
    \includegraphics[width=\textwidth]{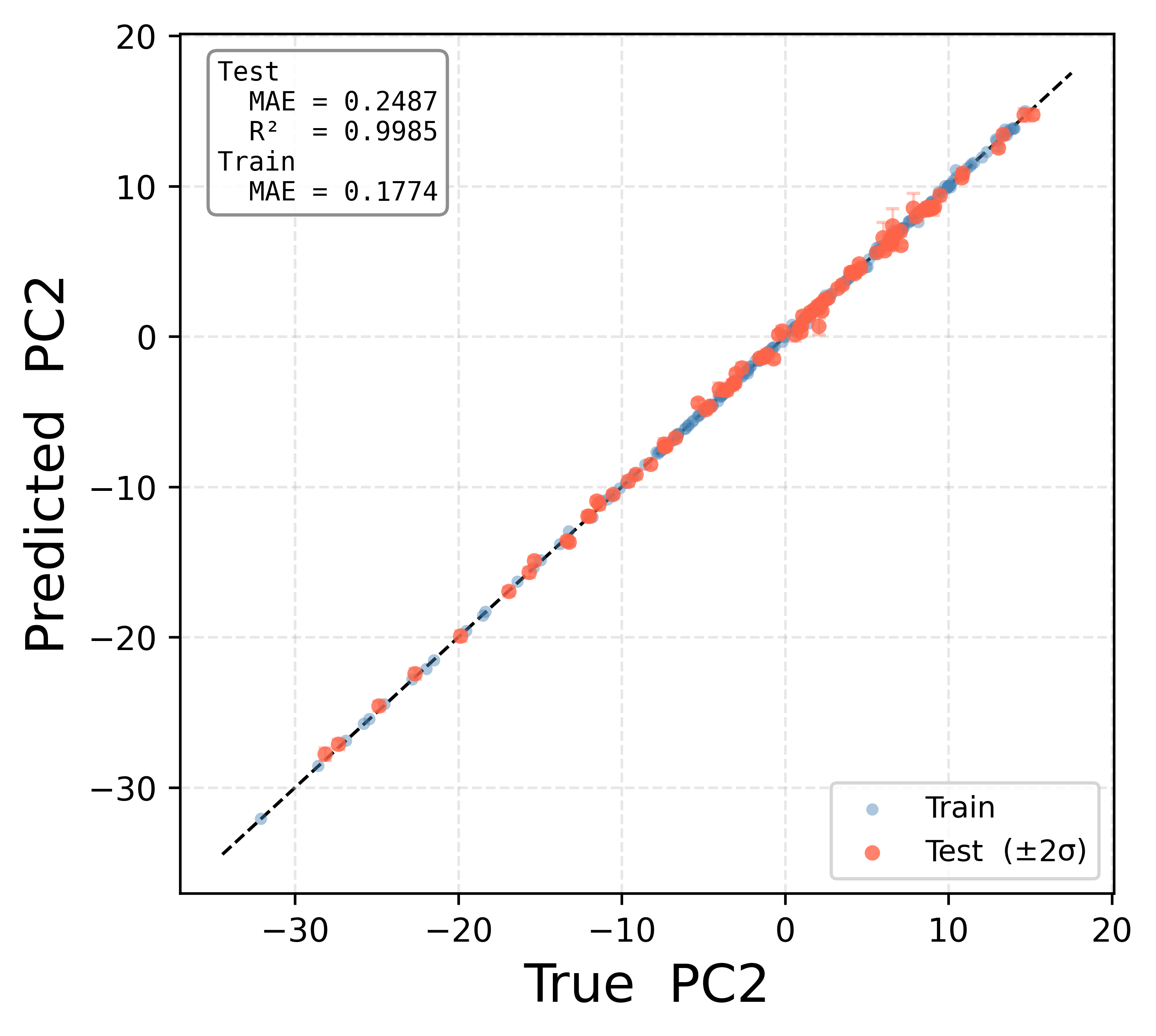}
    \caption{}
    \label{fig:parity_pc2}
\end{subfigure}
\hfill
\begin{subfigure}[b]{0.45\textwidth}
    \includegraphics[width=\textwidth]{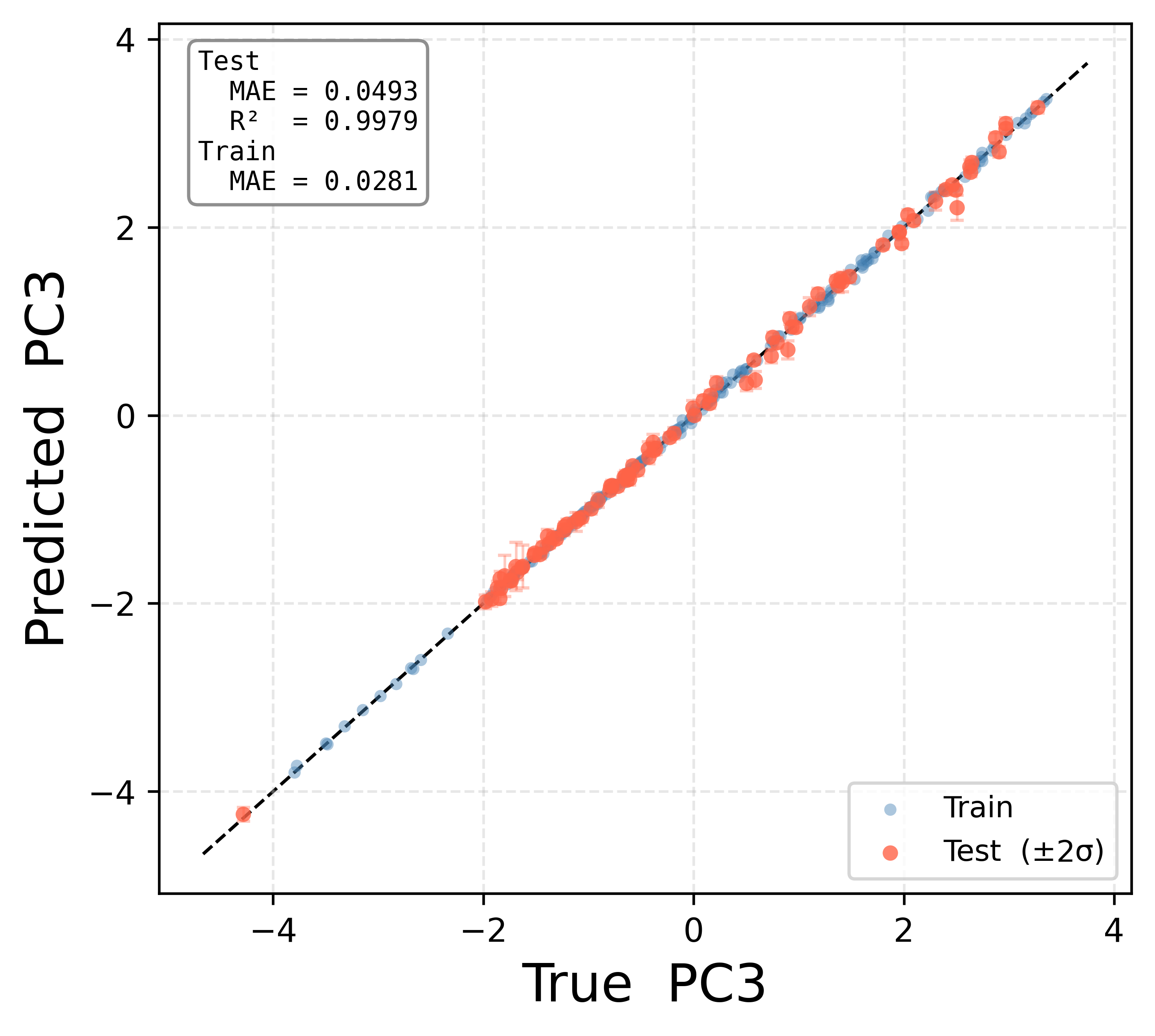}
    \caption{}
    \label{fig:parity_pc3}
\end{subfigure}
\caption{Parity plots of GP-predicted versus FE-extracted values for all five
retained features on the 30\% test set. The dashed line indicates the ideal
1:1 relation. MAE values are reported for each feature.}
\label{fig:feature_parity}
\end{figure*}

The GP predicts the extracted features with small errors throughout the test
set. We quantify the prediction error for each feature using
the mean absolute error
\begin{equation}
    \mathrm{MAE}_k =
    \frac{1}{N_\mathrm{test}}
    \sum_{i=1}^{N_\mathrm{test}}
    \bigl|y_k^{(i)} - \hat{y}_k^{(i)}\bigr|,
    \label{eq:feature_mae}
\end{equation}
where $y_k^{(i)}$ is feature $k$ for entry $i$ in the test set,
$\hat{y}_k^{(i)}$ is the GP predictive mean, and $N_\mathrm{test}=90$.
Because the features have different units and scales, their MAE values are
interpreted separately rather than compared directly across features.
The parity plots in Figure~\ref{fig:feature_parity} show that the predicted
features follow the FE-extracted values for both the $F$--$D$ and
DIC-derived quantities, with no systematic bias apparent over the sampled
parameter range.

\begin{figure*}[htbp]
\centering
\begin{subfigure}[b]{0.78\textwidth}
    \includegraphics[width=\textwidth]{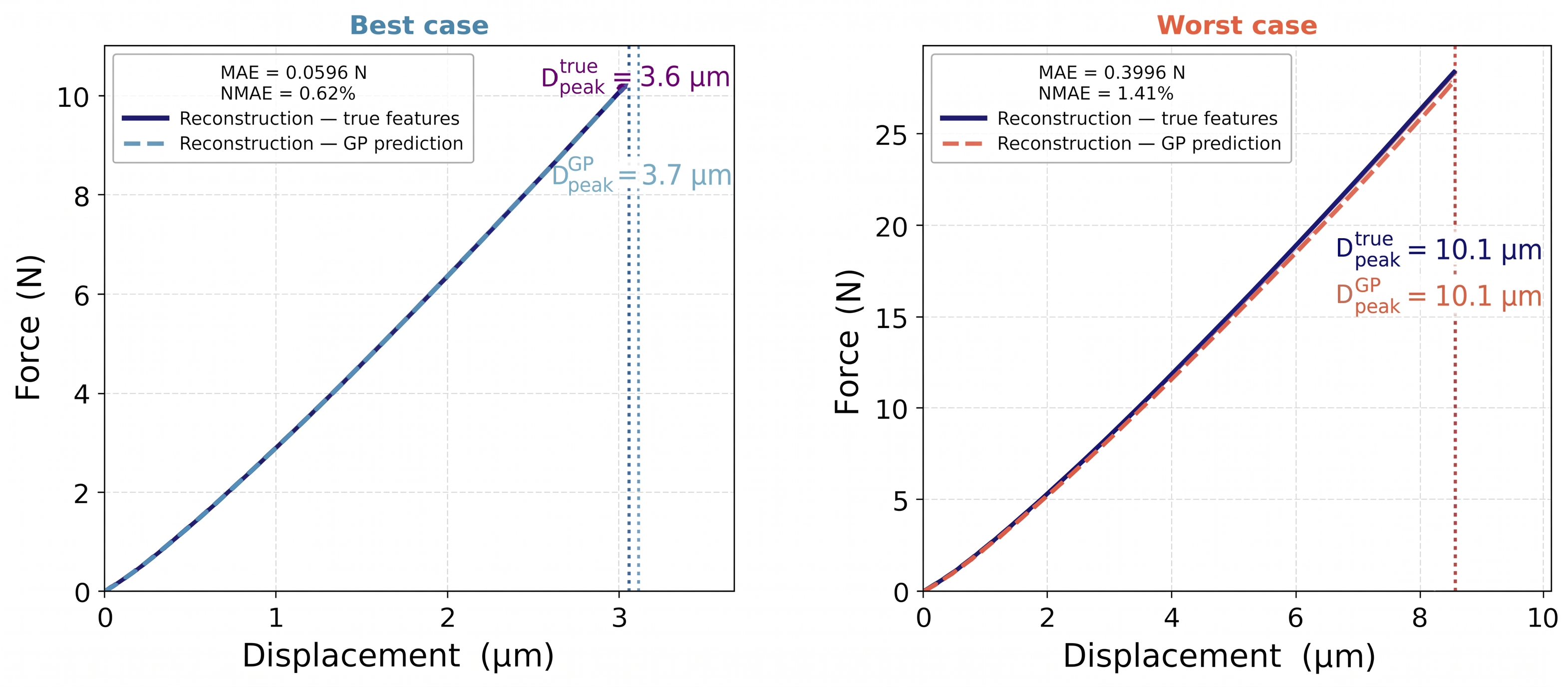}
    \caption{}
    \label{fig:gp_recon_curves}
\end{subfigure}
\hfill
\begin{subfigure}[b]{0.45\textwidth}
    \includegraphics[width=\textwidth]{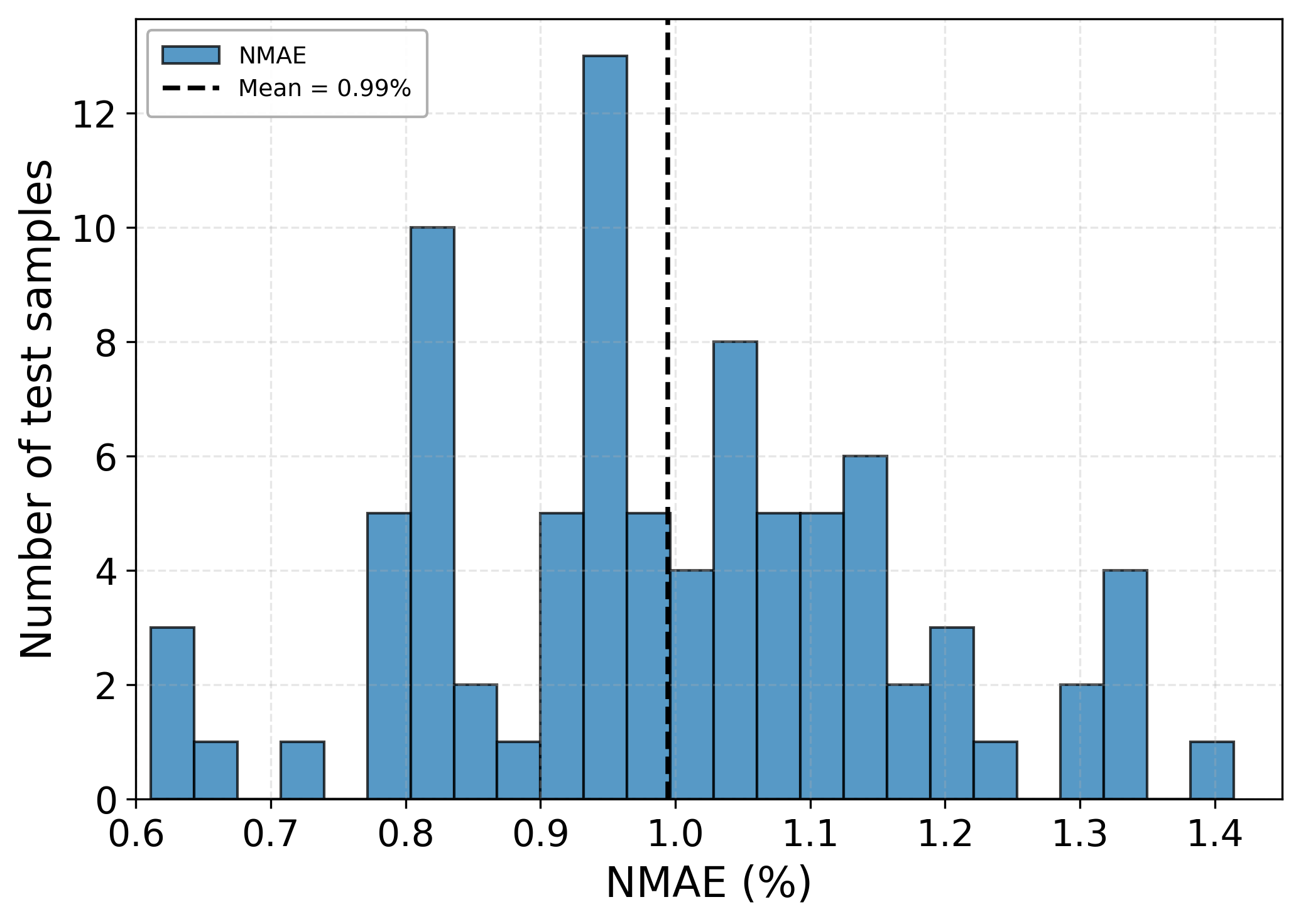}
    \caption{}
    \label{fig:nmae_dist}
\end{subfigure}
\caption{GP surrogate $F$--$D$ reconstruction quality on the 30\%
test set.
(a)~Best-case (left) and worst-case (right) reconstructions: the
predicted curve (dashed) agrees closely with the FE output
(solid) in the best case; the worst-case NMAE reaches 1.41\% with
the largest discrepancy near $D_\mathrm{peak}$.
(b)~Distribution of NMAE across the test set; the mean
NMAE is 0.99\%.}
\label{fig:gp_recon}
\end{figure*}

For each entry in the test set, we quantify the $F$--$D$
reconstruction error using the normalized mean absolute error
\begin{equation}
    \mathrm{NMAE}^{(i)}
    =
    \frac{
        \displaystyle \frac{1}{N_D}
        \sum_{j=1}^{N_D}
        \left|F_j^{(i)}-\widehat{F}_j^{(i)}\right|
    }{
        \overline{F}^{(i)}
    }
    \times 100\%,
    \qquad
    \overline{F}^{(i)}
    =
    \frac{1}{N_D}
    \sum_{j=1}^{N_D}F_j^{(i)},
    \label{eq:nmae}
\end{equation}
where $N_D$ is the number of displacement points,
$F_j^{(i)}$ is the FE-simulated force, and
$\widehat{F}_j^{(i)}$ is the GP-reconstructed force at displacement point
$j$. Figure~\ref{fig:gp_recon}a shows representative reconstructions of
the $F$--$D$ curve. In the best case, the predicted curve agrees with the FE
output at plotting resolution. In the worst case, the NMAE is 1.41\%, with
the largest discrepancy near $D_\mathrm{peak}$, where small feature
errors have the strongest effect on the reconstructed curve. Across the
test set, the mean NMAE is 0.99\%
(Figure~\ref{fig:gp_recon}b), indicating close agreement between the
GP-reconstructed and FE-simulated $F$--$D$ curves.

\begin{figure}[htbp]
\centering
\includegraphics[width=\columnwidth]{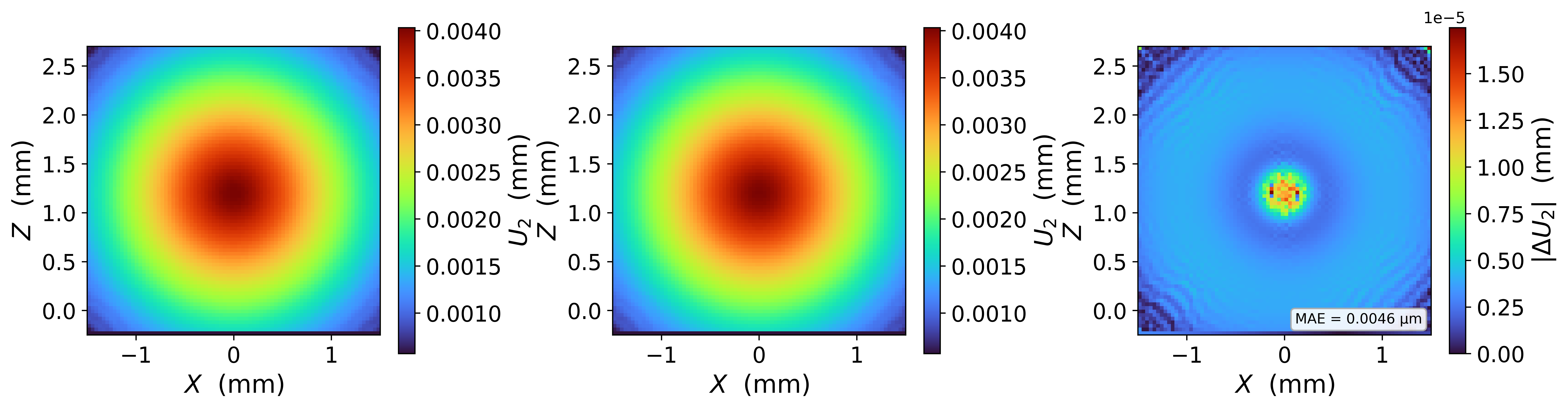}
\caption{Worst-case GP surrogate reconstruction of the DIC
displacement field $w(x,y)$ on the test set.
(Left) FE field; (center) GP-predicted field reconstructed from
predicted PC scores; (right) pointwise residual.  The residual is
spatially confined to the central punch contact zone with
$\mathrm{MAE}_w = 0.0046~\mu\mathrm{m}$, less than 0.2\% of the peak
field value.}
\label{fig:dic_recon}
\end{figure}

The same conclusion holds for the spatial DIC field. Even for the
worst-performing entry in the test set, the GP-predicted field reconstructed
from the predicted PC scores reproduces the central deflection dome and the
decay toward the supported region (Figure~\ref{fig:dic_recon}). The residual is
concentrated near the punch contact zone, where the displacement gradient is
largest, and the spatial mean absolute error is
$\mathrm{MAE}_w=0.0046~\mu\mathrm{m}$, less than 0.2\% of the peak field value.
The three-component PCA--GP representation therefore retains the dominant
spatial variation of the selected DIC displacement field.

These surrogate checks establish the computational basis for the posterior
comparison. The GP reproduces the five-feature FE mapping with small
errors across the test set in both scalar and field-derived quantities. We
therefore next evaluate the trained CFM posterior estimator using experimental
features extracted from the representative SPT--DIC measurement.

\subsection{Posterior inference using force--displacement and DIC features}
\label{sec:posterior_inference}

The DIC displacement field substantially contracts the posterior relative to
$F$--$D$ conditioning alone. We isolate this effect by comparing
two inference cases under the same prior bounds and GP surrogate. The first
uses only the $F$--$D$ features,
$(A,D_\mathrm{peak})$, and the measured specimen thickness. The
second adds the three DIC PC scores at $D_\mathrm{peak}$.

\begin{figure*}[htbp]
\centering
\includegraphics[width=0.65\textwidth]{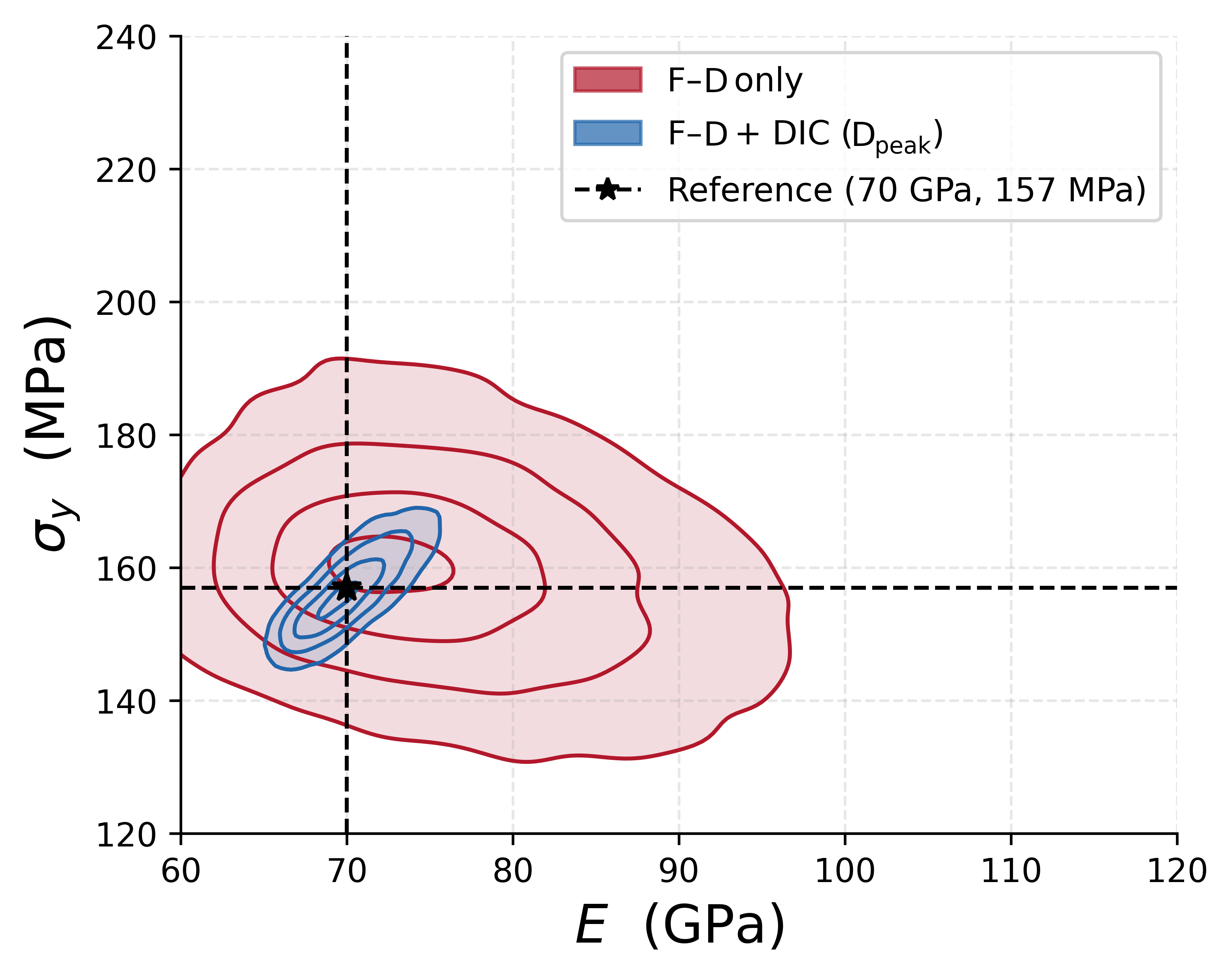}
\caption{Joint posterior distributions for Young's modulus $E$ and yield
strength $\sigma_y$ under two conditioning cases: $F$--$D$
features only and combined $F$--$D$ plus DIC features at
$D_\mathrm{peak}$. Contours show posterior density levels from
20\,000 CFM samples. The dashed lines and star mark
the independent tensile
reference values, $E=70$~GPa and $\sigma_y=157$~MPa. The DIC features reduce
the posterior spread and concentrate the posterior closer to the reference
values while retaining a positive $E$--$\sigma_y$ correlation.}
\label{fig:joint_posterior}
\end{figure*}

The force--displacement features alone produce a broad posterior in the
$(E,\sigma_y)$ plane (Figure~\ref{fig:joint_posterior}). This posterior
spread reflects the mechanical coupling of elastic stiffness and yield
strength in the early SPT response: different parameter combinations can
produce similar global $F$--$D$ features.

Adding the DIC displacement-field features substantially narrows the
posterior distribution. The $F$--$D$+DIC posterior is concentrated near the
independent tensile reference values, while retaining a positive
$E$--$\sigma_y$ correlation over a smaller region of parameter space. This
posterior contraction indicates that the spatial displacement field adds
information that is not captured by the scalar $F$--$D$ features.
The residual positive correlation is consistent with the mechanical coupling
between elastic stiffness and yield strength in the early SPT response.

\begin{figure}[htbp]
\centering
\includegraphics[width=\columnwidth]{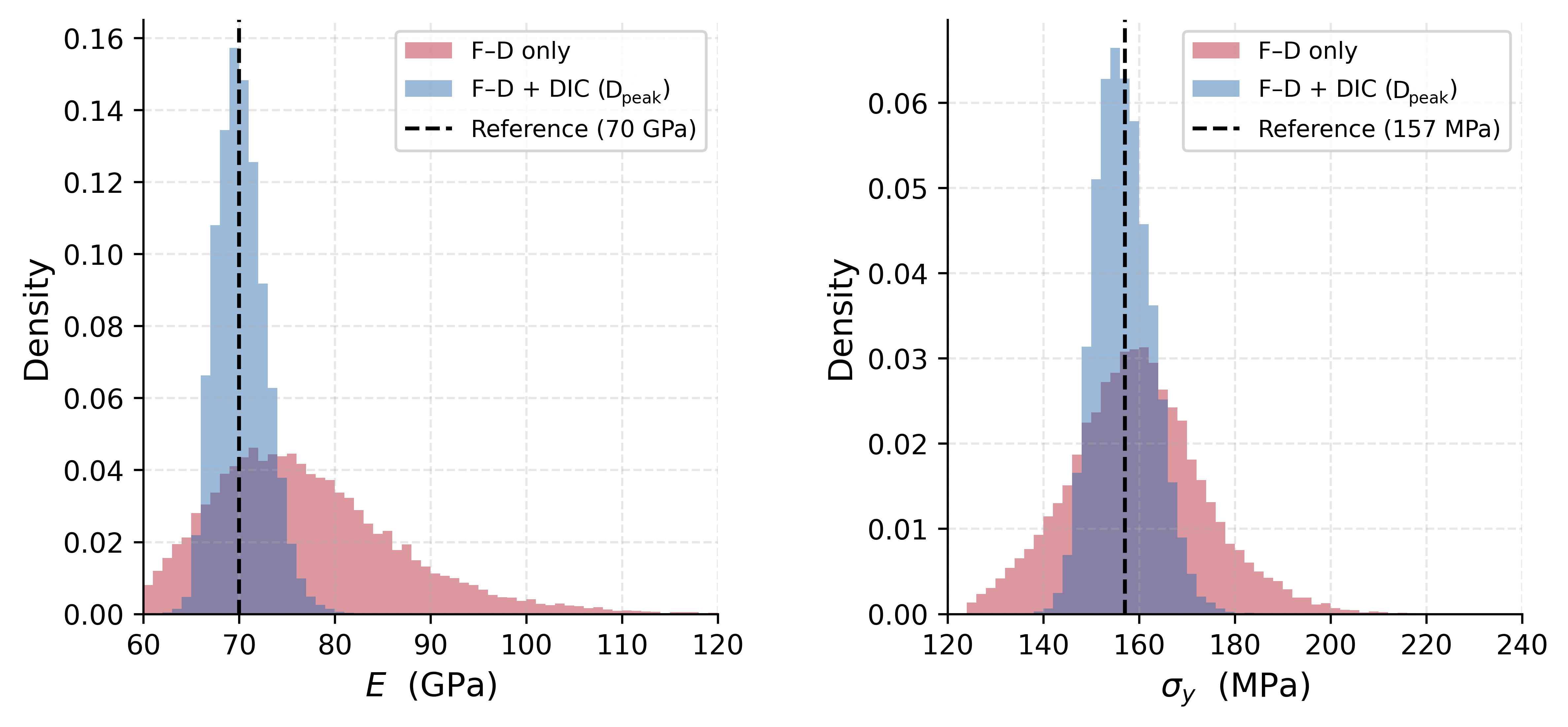}
\caption{Marginal posterior distributions for $E$ (left) and
$\sigma_y$ (right) using $F$--$D$ features only and combined
$F$--$D$ plus DIC features. Reference values from independent
tensile tests are indicated by dashed vertical lines. The DIC features
contract the marginal posteriors for both parameters.}
\label{fig:posterior_marginals}
\end{figure}

The marginal posteriors quantify this contraction
(Figure~\ref{fig:posterior_marginals} and
Table~\ref{tab:posterior_summary}). In the $F$--$D$-only case,
the 95\% highest posterior density (HPD) credible interval for $E$ is
$[61.2,\,101.0]$~GPa, with a width of 39.8~GPa, and the interval for
$\sigma_y$ is $[132.7,\,189.7]$~MPa, with a width of 57.0~MPa. Both
intervals include the tensile reference values, but the posterior remains
broad, particularly for $E$. The lower limit of the $E$ interval lies close
to the prior bound of 60~GPa, indicating that the $F$--$D$
features provide limited constraint on the lower tail of the posterior. The
posterior medians are 75.57~GPa for $E$ and 159.29~MPa for $\sigma_y$.

When the DIC features are included, the 95\% HPD credible interval contracts to
$[65.9,\,75.7]$~GPa for $E$ and $[146.4,\,169.1]$~MPa for $\sigma_y$,
corresponding to interval widths of 9.8~GPa and 22.7~MPa, respectively.
Relative to the $F$--$D$-only posterior, the interval width is
reduced by a factor of 4.1 for $E$ and 2.5 for $\sigma_y$. The corresponding
posterior medians, 70.03~GPa and 156.34~MPa, are close to the independent
tensile reference values. This
factor-of-4.1 contraction in $E$ and factor-of-2.5 contraction in
$\sigma_y$, combined with the shift of the posterior medians toward the
tensile reference values, is consistent with improved parameter
identifiability when the global $F$--$D$ response is supplemented
by the spatial displacement field.

\begin{table}[htbp]
\centering
\caption{Posterior summary for the two inference scenarios.
Bounds are 95\% highest posterior density (HPD) credible interval limits;
the width is the interval span.}
\label{tab:posterior_summary}
\begin{tabular}{lcccccc}
\toprule
 & \multicolumn{3}{c}{$E$ (GPa)} & \multicolumn{3}{c}{$\sigma_y$ (MPa)} \\
\cmidrule(lr){2-4}\cmidrule(lr){5-7}
Scenario & $\tilde{E}$ & 95\% HPD & Width & $\tilde{\sigma}_y$ & 95\% HPD & Width \\
\midrule
$F$--$D$ only  & 75.57 & [61.2,\;101.0] & 39.8 & 159.29 & [132.7,\;189.7] & 57.0 \\
$F$--$D$+DIC  & 70.03 & [65.9,\;75.7] & 9.8 & 156.34 & [146.4,\;169.1] & 22.7 \\
Tensile reference & 70.0 & -- & -- & 157.0 & -- & -- \\
\bottomrule
\end{tabular}
\end{table}

\subsection{Posterior calibration diagnostics}
\label{sec:posterior_diagnostics}

Narrow posterior intervals are useful only if they contain the true parameter
values at the stated probability. We therefore assess the calibration of the
$F$--$D$+DIC estimator using the test set ($N_\mathrm{test}=90$), which was
excluded from GP and CFM training. The values of $E$ and $\sigma_y$ are known
for every entry in the test set. Synthetic features were generated for each
entry, and 2000 posterior samples were drawn. Calibration was then assessed
using empirical coverage and rank histograms, as described in
Appendix~\ref{app:sbc}.

\begin{figure*}[htbp]
\centering
\includegraphics[width=0.9\textwidth]{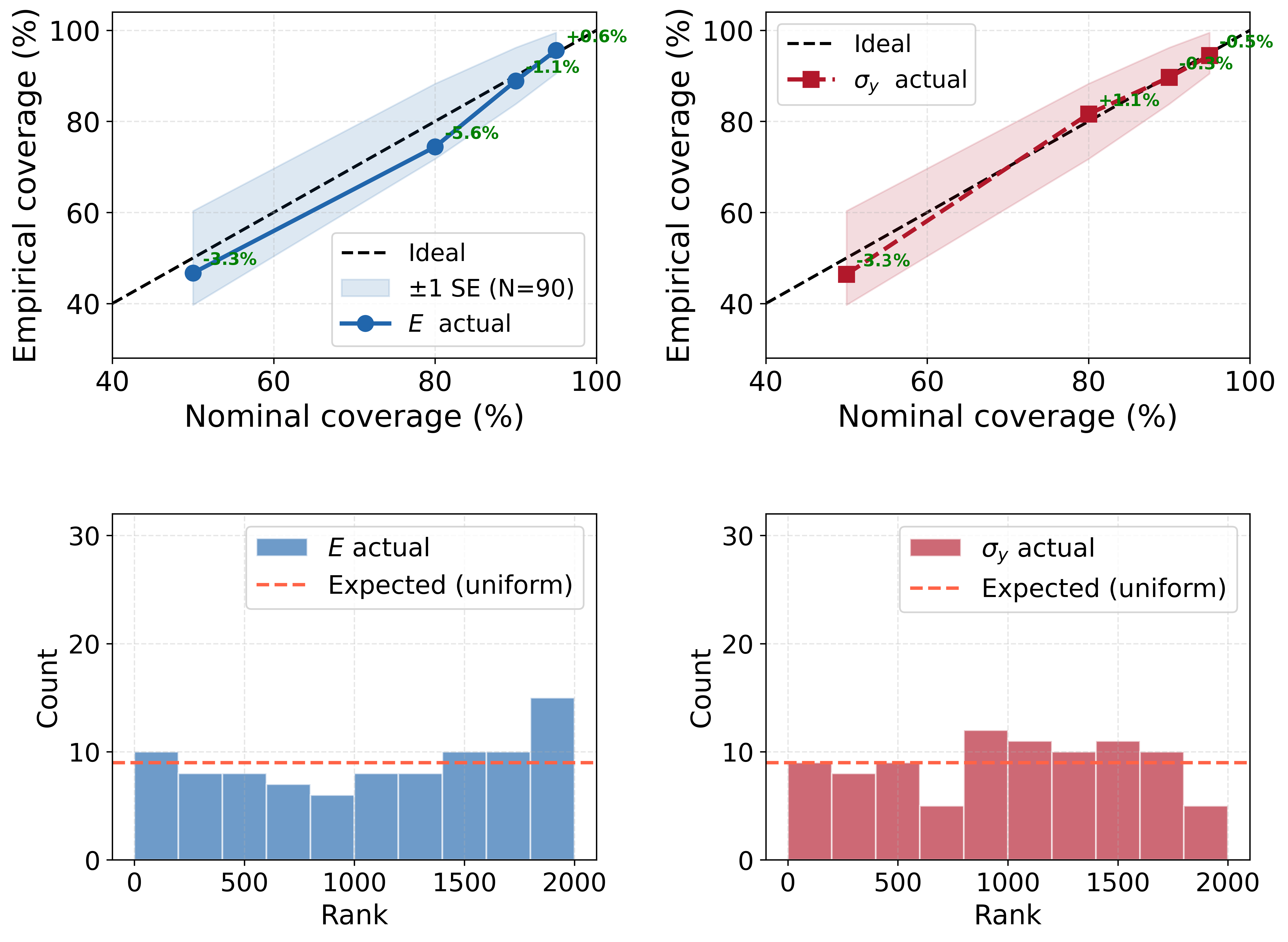}
\caption{Posterior calibration diagnostics for the $F$--$D$+DIC inference
scenario, evaluated using synthetic observations generated at the 90
parameter--thickness combinations in the test set, with 2000 CFM posterior
samples for each entry. Top row: empirical coverage versus nominal
coverage for $E$ (left) and $\sigma_y$ (right); the dashed line denotes ideal
calibration and shaded bands show one standard error for $N=90$. Bottom row:
rank histograms for $E$ and $\sigma_y$; the dashed horizontal line marks the
expected uniform frequency.}
\label{fig:coverage_diagnostics}
\end{figure*}

Table~\ref{tab:coverage_summary} and Figure~\ref{fig:coverage_diagnostics}
summarize the coverage results. At the 90\% and 95\% levels, the empirical
coverage is close to the corresponding nominal value for both parameters,
with deviations no greater than 1.1 percentage points. Larger departures
occur at the lower coverage levels, particularly for $E$ at 80\%. This point
lies slightly outside the one-standard-error band in
Figure~\ref{fig:coverage_diagnostics}; all other reported coverage values lie
within their corresponding bands. The results therefore indicate mild
overconfidence in the central part of the posterior for $E$, while the
high-coverage intervals used to report the experimental results remain close
to their nominal levels. A parallel calibration assessment of the
$F$--$D$-only baseline is not included here.

\begin{table}[htbp]
\centering
\caption{Empirical coverage of the $F$--$D$+DIC posterior at four nominal
levels, evaluated using synthetic observations generated at the 90
parameter--thickness combinations in the test set, with 2000 CFM posterior
samples for each entry. $\Delta$ denotes empirical minus nominal coverage in
percentage points.}
\label{tab:coverage_summary}
\begin{tabular}{ccccc}
\toprule
Nominal (\%) & $E$ empirical (\%) & $\Delta_E$ (pp) &
$\sigma_y$ empirical (\%) & $\Delta_{\sigma_y}$ (pp) \\
\midrule
50 & 46.7 & $-3.3$ & 46.7 & $-3.3$ \\
80 & 74.4 & $-5.6$ & 81.1 & $+1.1$ \\
90 & 88.9 & $-1.1$ & 90.0 & $-0.3$ \\
95 & 95.6 & $+0.6$ & 95.6 & $+0.5$ \\
\bottomrule
\end{tabular}
\end{table}

The rank histograms provide a complementary check of the full posterior
distribution. A calibrated scalar posterior produces ranks that are
approximately uniform when the true parameter value is inserted among
posterior samples. The histograms show some bin-to-bin variation but no clear
U-shaped pattern, which would indicate overconfidence, or mound-shaped
pattern, which would indicate underconfidence. These diagnostics assess the
calibration of the CFM estimator under the surrogate-based observation
generator using synthetic observations generated for the test set; they do
not assess GP bias relative to FE outputs or validate the FE model against the
experimental system.

\subsection{Posterior predictive consistency}
\label{sec:posterior_predictive}

\begin{figure*}[htbp]
\centering
\begin{subfigure}[b]{0.48\textwidth}
    \includegraphics[width=\textwidth]{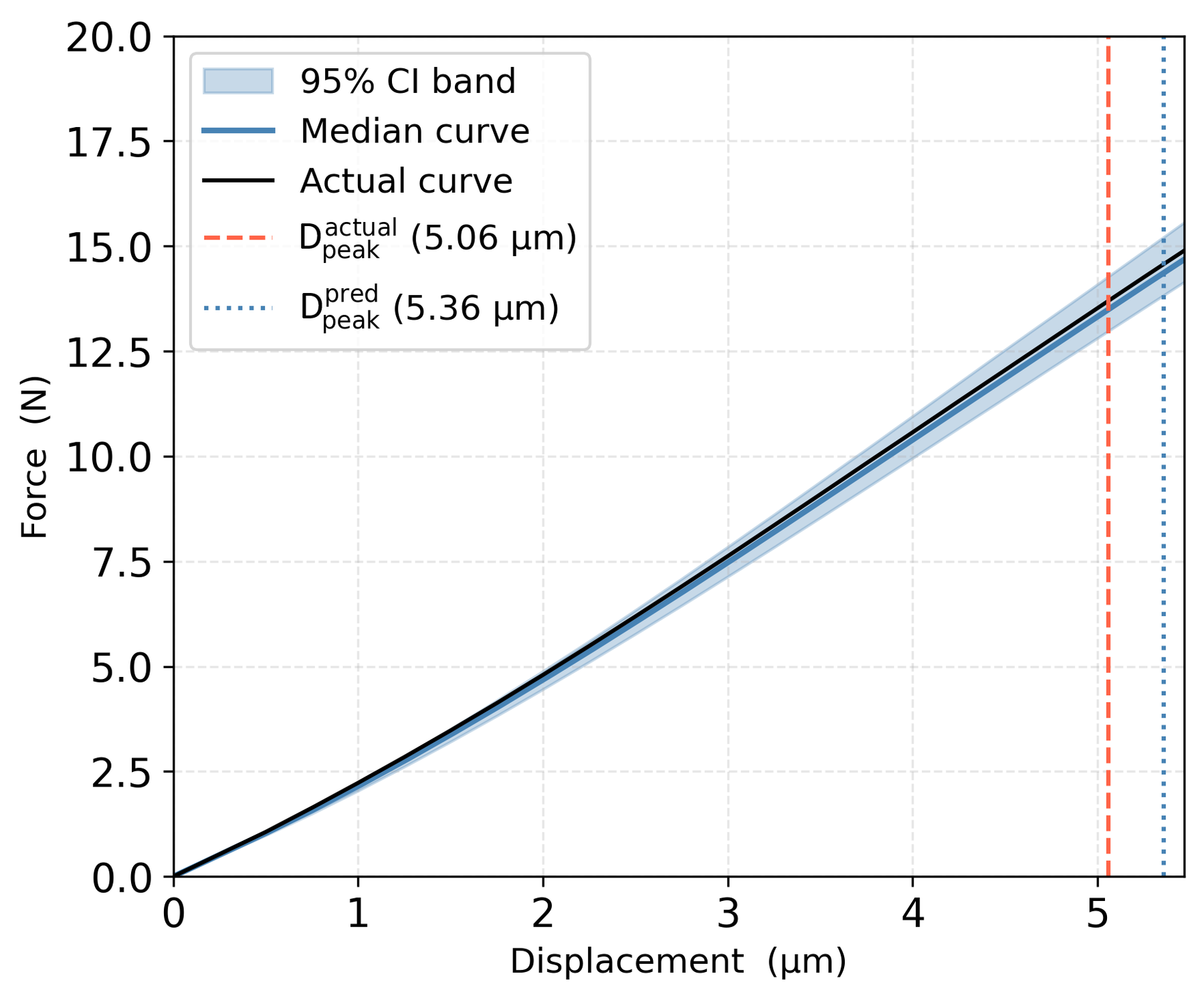}
    \caption{$F$--$D$ only}
    \label{fig:fd_ppc_nomic}
\end{subfigure}
\hfill
\begin{subfigure}[b]{0.48\textwidth}
    \includegraphics[width=\textwidth]{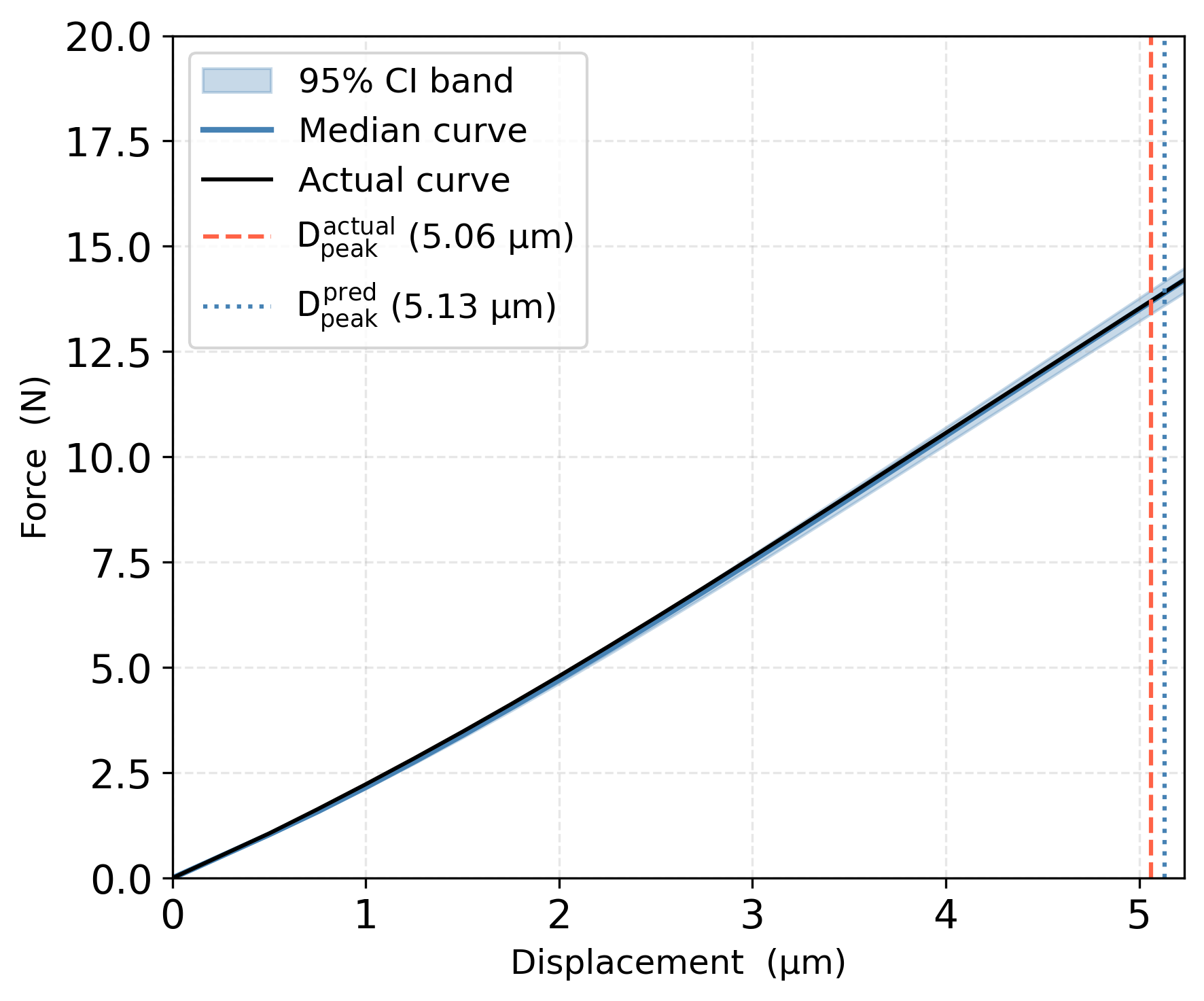}
    \caption{$F$--$D$ and DIC}
    \label{fig:fd_ppc_dic}
\end{subfigure}
\caption{Posterior predictive checks for the $F$--$D$ curve under the
two inference scenarios. The 95\% credible band (shaded) and median
curve (solid blue) are obtained by propagating 20\,000 posterior
samples through the GP surrogate; the measured curve is
shown in black. Red and blue dashed vertical lines mark the measured
and predicted $D_\mathrm{peak}$, respectively.
(a)~$F$--$D$ alone. (b)~$F$--$D$ combined with DIC, for which the
$D_\mathrm{peak}$ offset is reduced from 0.30 to 0.07~$\mu$m.}
\label{fig:fd_ppc}
\end{figure*}

We next examine whether the posterior contraction obtained by adding DIC
preserves agreement with the measured $F$--$D$ response. When
posterior samples are propagated through the GP surrogate, the resulting
values of $A$ are used to reconstruct the predicted $F$--$D$ curves
using $F=AD^{1.15}$ (Eq.~\ref{eq:powerlaw_fd}). The median predicted curves
follow the measured curve for both conditioning cases
(Figure~\ref{fig:fd_ppc}). In the $F$--$D$-only case, the
posterior-predicted $D_\mathrm{peak}$ is offset from the measured value by
0.30~$\mu$m (Table~\ref{tab:ppc_summary}). Inclusion of the DIC features
decreases this offset to 0.07~$\mu$m. The $F$--$D$+DIC posterior therefore
remains consistent with the global response while reducing uncertainty in the
inferred parameters.

\begin{table}[htbp]
\centering
\caption{$F$--$D$ posterior predictive check: predicted versus measured
$D_\mathrm{peak}$ under the two
inference scenarios. The measured value is 5.06~$\mu$m.}
\label{tab:ppc_summary}
\begin{tabular}{lccc}
\toprule
Scenario & Predicted ($\mu$m) & Measured ($\mu$m) & Offset ($\mu$m) \\
\midrule
$F$--$D$ only & 5.36 & 5.06 & 0.30 \\
$F$--$D$+DIC & 5.13 & 5.06 & 0.07 \\
\bottomrule
\end{tabular}
\end{table}

We also examine whether the posterior mean obtained from the
$F$--$D$+DIC inference reproduces the measured spatial displacement pattern.
The posterior-mean displacement field is reconstructed as
$\hat{\mathbf{w}}=\bar{\mathbf{w}}+\mathbf{U}\bar{\mathbf{s}}$, where
$\bar{\mathbf{w}}$ is the mean displacement field of the FE training set,
$\mathbf{U}$ is the PCA basis trained on the FE dataset as described in
Section~\ref{sec:dic_features}, and
$\bar{\mathbf{s}} =
(\mathbb{E}[\mathrm{PC}_1],\mathbb{E}[\mathrm{PC}_2],
\mathbb{E}[\mathrm{PC}_3])^\top$ is the posterior mean PC-score vector.

\begin{figure}[htbp]
\centering
\includegraphics[width=\columnwidth]{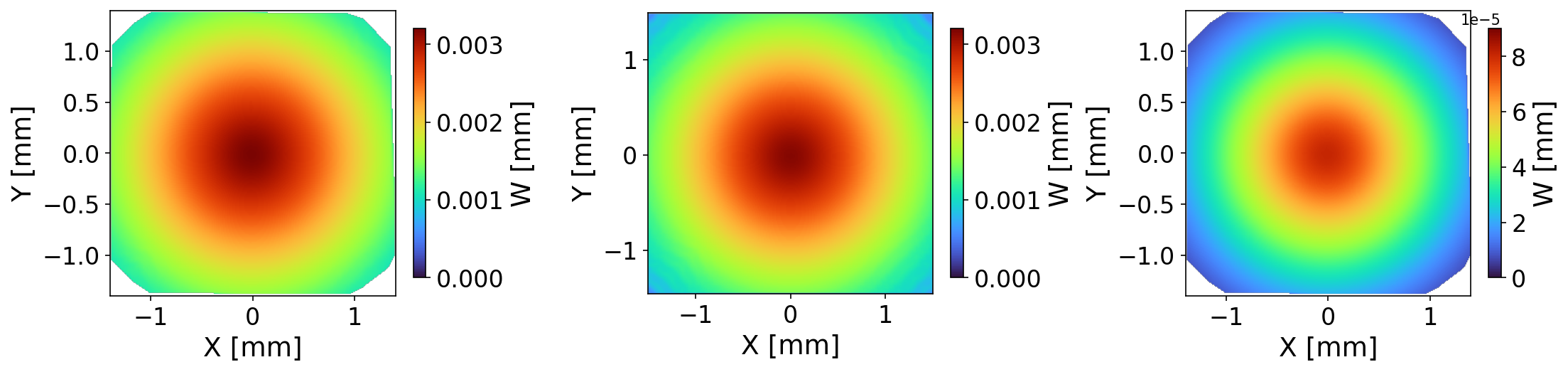}
\caption{Comparison of the experimentally measured DIC out-of-plane
displacement field $w_\mathrm{exp}(x,y)$ with the posterior-mean
reconstructed field $\hat{w}(x,y)$
under combined $F$--$D$+DIC inference, evaluated at the
$D_\mathrm{peak}$ loading frame.
(Left)~Experimental field; (center)~posterior-mean reconstructed field;
(right)~pointwise residual $w_\mathrm{exp} - \hat{w}$.  Both fields
share a peak displacement of $\approx 0.003$~mm and reproduce the
radially symmetric deflection dome and die-opening boundary.  The
maximum absolute residual is $8\times10^{-5}$~mm ($\approx 2.7\%$ of
the peak value), concentrated beneath the center of the punch.}
\label{fig:dic_exp_pred}
\end{figure}

The experimental and reconstructed fields have the same peak
out-of-plane displacement scale, approximately 0.003~mm, and both exhibit the
die-opening boundary, the central deflection dome, and the decay toward the
clamped edges (Figure~\ref{fig:dic_exp_pred}). The residual is
largest in the central region beneath the punch, where displacement gradients are
steepest. Its maximum absolute value is $8\times10^{-5}$~mm, approximately
2.7\% of the peak field value. This agreement supports the consistency of
the inferred parameters with both the global $F$--$D$ response and
the measured DIC displacement field.

Together, the $F$--$D$ posterior predictive check and the
posterior-mean DIC-field comparison show that the multimodal posterior
preserves agreement with both measured response modalities. This consistency
supports the use of DIC information at $D_\mathrm{peak}$ as a complementary
constraint for two-parameter SPT-based inference.

\section{Conclusions}
\label{sec:conclusions}

This study developed an amortized, likelihood-free Bayesian framework for
jointly inferring Young's modulus $E$ and yield strength $\sigma_y$ from
multimodal SPT measurements. Gaussian process surrogates were used to emulate
the relationship between the FE model inputs and the extracted response
features, and a CFM estimator was trained to sample the corresponding
conditional posterior. The framework was demonstrated using the early
force--displacement response and the bottom-surface DIC displacement field
measured for an AA6111-T4 specimen. Independent tensile measurements were
used only as external reference values and were not supplied to the inference.

The force--displacement features alone produced broad, correlated posteriors,
consistent with the limited parameter discrimination of the global response.
Adding the three DIC PC scores at $D_\mathrm{peak}$ reduced the 95\% HPD
interval width from 39.8 to 9.8~GPa for $E$ and from 57.0 to 22.7~MPa for
$\sigma_y$, corresponding to contraction factors of 4.1 and 2.5,
respectively. The multimodal posterior medians, 70.03~GPa and 156.34~MPa,
were also close to the independent tensile reference values of 70~GPa and
157~MPa. These results show that the bottom-surface displacement field
provides information that reduces the uncertainty remaining when only the
global force--displacement response is used.

The GP surrogate reproduced the responses in the test set with small errors.
The mean NMAE of the reconstructed force--displacement curves was 0.99\%, and the
worst-case reconstructed displacement field had a spatial MAE of
$0.0046~\mu\mathrm{m}$, less than 0.2\% of its peak value. This accuracy
was obtained using 210 FE simulations for surrogate fitting and 90 for
testing; after this assessment, all 300 simulations were used to fit the
surrogate for experimental inference. By comparison, a recent GP-based study
of SPT force--displacement responses employed 850 FE
simulations~\cite{courtright2025htspt}. Although that study considered a
broader constitutive input space and is therefore not directly comparable,
the present results show that the compact five-feature representation
enabled accurate surrogate modeling of the three-input map using a
comparatively modest FE dataset. A
force--displacement posterior predictive check and a posterior-mean DIC-field
comparison showed that the multimodal posterior remained consistent with
both measured response modalities. Adding DIC reduced the posterior-predicted
$D_\mathrm{peak}$ offset from 0.30 to 0.07~$\mu\mathrm{m}$, while the maximum
absolute residual in the posterior-mean displacement field was approximately
2.7\% of the measured peak displacement. Surrogate accuracy and calibration
were assessed using the FE test set, whereas response consistency was
evaluated against the measured SPT--DIC data. These checks support the
internal consistency of the GP--CFM inference pipeline but do not constitute
an independent validation of the FE model.

Once trained, the CFM estimator generates posterior samples for a new
specimen by integrating the learned transport field, without constructing an
experimental likelihood or initiating a new MCMC chain. The computational
cost is therefore concentrated in the initial FE simulation and training
stages, while subsequent inference reuses the trained models.

The present demonstration was deliberately limited to a two-parameter EPP
model and a single DIC field evaluated at $D_\mathrm{peak}$. Extending the
analysis to hardening or damage models will require later-stage measurements
that are informative about the additional constitutive parameters, together
with an expanded FE dataset and renewed training and validation of the
surrogate and posterior estimator.
Within the scope examined here, the results demonstrate that incorporating
bottom-surface DIC information can substantially improve uncertainty-quantified
identification of $E$ and $\sigma_y$ from the early SPT response.

\section*{CRediT authorship contribution statement}
\textbf{Mohammad Ali Seyed Mahmoud:} Writing--original draft, Validation,
Visualization, Methodology, Software, Investigation, Formal analysis, Data
curation, Conceptualization. \textbf{Aditya Venkatraman:} Writing--review
\& editing, Formal analysis. \textbf{Raj Mahat:} Investigation.
\textbf{Samantha Mitra:} Investigation. \textbf{Surya R. Kalidindi:}
Writing--review \& editing, Supervision, Visualization, Funding acquisition,
Conceptualization.

\section*{Acknowledgements}
This work was supported by the National Science Foundation (NSF) under grant
number 2221104 and by the Army Research Laboratory under Cooperative
Agreement Number W911NF-22-2-0106. The views and conclusions contained in
this document are those of the authors and should not be interpreted as
representing the official policies, either expressed or implied, of the Army
Research Laboratory or the U.S. Government. The U.S. Government is
authorized to reproduce and distribute reprints for Government purposes
notwithstanding any copyright notation herein. The authors gratefully
acknowledge the support of the George W. Woodruff School of Mechanical
Engineering for providing the necessary facilities and resources.

\section*{Declaration of competing interest}
The authors declare that they have no known competing financial interests or
personal relationships that could have appeared to influence the work
reported in this paper.

\section*{Data availability}
Data will be made available on request.

\appendix
\section{CFM implementation details}
\label{app:cfm_details}

The velocity network uses three hidden layers with 64 units per layer and
SiLU activations. The network is trained for 2000 epochs using the Adam
optimizer~\cite{kingma2015adam}, a batch size of 128, a base learning rate
of $10^{-3}$, and a cosine annealing learning-rate scheduler. An exponential
moving average (EMA) of the network weights is maintained to improve
trajectory smoothness during inference. The hyperparameters are summarized
in Table~\ref{tab:cfm_params}.

\begin{table}[htbp]
\centering
\caption{Hyperparameters of the Conditional Flow Matching model.}
\label{tab:cfm_params}
\begin{tabular}{lll}
\toprule
\textbf{Category} & \textbf{Parameter} & \textbf{Value} \\
\midrule
\textit{Architecture} & Hidden layers & $3 \times 64$ \\
& Activation function & SiLU \\
& Time embedding & Sinusoidal \\
& Conditioning injection & Concatenation \\
\midrule
\textit{Training} & Epochs & 2000 \\
& Batch size & 128 \\
& Base learning rate & $10^{-3}$ \\
& Learning-rate scheduler & Cosine annealing \\
& Optimizer & Adam \\
& EMA decay rate & 0.999 \\
& GP-augmented samples & 1000 per epoch \\
\midrule
\textit{Inference} & ODE solver & RK4 \\
& ODE steps & 20 \\
& Posterior samples & 20\,000 \\
\bottomrule
\end{tabular}
\end{table}

Algorithm~\ref{alg:cfm_sampling} summarizes the amortized posterior sampling
procedure. The final step maps the transported samples from normalized
coordinates back to physical parameter space.

\begin{algorithm}[htbp]
\caption{Amortized posterior sampling with the trained CFM model}
\label{alg:cfm_sampling}
\DontPrintSemicolon
\SetKwInOut{Input}{Input}
\SetKwInOut{Output}{Output}

\Input{trained velocity field $\mathbf{v}_\phi$; conditioning vector
$\mathbf{c}_\mathrm{obs}$; posterior sample count $N_\mathrm{post}$;
ODE step count $M$; parameter normalization statistics
$(\bm{\mu}_\theta,\bm{\sigma}_\theta)$}
\Output{posterior samples
$\{\hat{\bm{\theta}}^{(s)}\}_{s=1}^{N_\mathrm{post}}$}

\BlankLine
$\Delta\tau \leftarrow 1/M$\;
Draw initial samples
$\mathbf{z}^{(s)}_0\sim\mathcal{N}(\mathbf{0},\mathbf{I})$,
$s=1,\ldots,N_\mathrm{post}$\;
\tcp*{All samples are propagated in parallel as a batched tensor}

\BlankLine
\For{$m=0,\ldots,M-1$}{
    $\tau_m \leftarrow m\Delta\tau$\;

    $\mathbf{k}_1^{(s)}
    \leftarrow
    \mathbf{v}_\phi\!\left(\mathbf{z}^{(s)}_m,\tau_m,
    \mathbf{c}_\mathrm{obs}\right)$\;

    $\mathbf{k}_2^{(s)}
    \leftarrow
    \mathbf{v}_\phi\!\left(\mathbf{z}^{(s)}_m
    +\frac{\Delta\tau}{2}\mathbf{k}_1^{(s)},
    \tau_m+\frac{\Delta\tau}{2},
    \mathbf{c}_\mathrm{obs}\right)$\;

    $\mathbf{k}_3^{(s)}
    \leftarrow
    \mathbf{v}_\phi\!\left(\mathbf{z}^{(s)}_m
    +\frac{\Delta\tau}{2}\mathbf{k}_2^{(s)},
    \tau_m+\frac{\Delta\tau}{2},
    \mathbf{c}_\mathrm{obs}\right)$\;

    $\mathbf{k}_4^{(s)}
    \leftarrow
    \mathbf{v}_\phi\!\left(\mathbf{z}^{(s)}_m
    +\Delta\tau\,\mathbf{k}_3^{(s)},
    \tau_m+\Delta\tau,
    \mathbf{c}_\mathrm{obs}\right)$\;

    $\mathbf{z}^{(s)}_{m+1}
    \leftarrow
    \mathbf{z}^{(s)}_m
    +
    \frac{\Delta\tau}{6}
    \left(
    \mathbf{k}_1^{(s)}
    +2\mathbf{k}_2^{(s)}
    +2\mathbf{k}_3^{(s)}
    +\mathbf{k}_4^{(s)}
    \right)$\;
}

\BlankLine
\For{$s=1,\ldots,N_\mathrm{post}$}{
    $\hat{\bm{\theta}}^{(s)}
    \leftarrow
    \bm{\mu}_\theta
    +
    \mathbf{z}^{(s)}_M\odot\bm{\sigma}_\theta$\;
}

\Return{$\{\hat{\bm{\theta}}^{(s)}\}_{s=1}^{N_\mathrm{post}}$}\;
\end{algorithm}

\section{Posterior predictive checks}
\label{app:ppc}

A posterior predictive check (PPC) tests whether the parameter values
inferred from the data are internally consistent with the observations used
for conditioning~\cite{gelman1996ppc}. The check does not require access to
ground-truth parameter values and can therefore be applied directly to
experimental data.

\textbf{Computation.} Given the experimental feature vector
$\mathbf{y}_\mathrm{obs}$ and the corresponding posterior samples
$\{\hat{\bm{\theta}}^{(s)}\}_{s=1}^{N_\mathrm{post}}$, each sample is
propagated through the stochastic observation model, combining the GP
predictive distribution with the feature-level noise, to obtain a replicated
feature vector. Writing $\mathbf{x}^{(s)}=(\hat{\bm{\theta}}^{(s)},t)$ for the
corresponding GP input,
\begin{equation}
    y_k^{\mathrm{rep},(s)}
    \sim
    \mathcal{N}\!\left(
    \hat{y}_k(\mathbf{x}^{(s)}),\;
    \sigma^2_{\mathrm{GP},k}(\mathbf{x}^{(s)})
    + \nu_k
    \right),
    \qquad s = 1,\ldots,N_\mathrm{post}.
    \label{eq:ppc}
\end{equation}
This produces a posterior predictive distribution of replicated features,
which can be compared directly with the observed feature vector.

\textbf{Interpretation.} If the inferred parameters are consistent with the
data, the observed feature vector $\mathbf{y}_\mathrm{obs}$ should fall
within the spread of the predicted distribution
$\{\mathbf{y}^{\mathrm{rep},(s)}\}$. Systematic offsets between the predicted
distribution and $\mathbf{y}_\mathrm{obs}$---for example, the observed
value lying outside the central 95\% of the predicted distribution for one
or more features---indicate residual model--data discrepancy. Common causes
include model form error (the constitutive model does not represent the true
material behavior), feature-extraction inconsistency between simulation and
experiment, or a likelihood specification that does not adequately
characterize measurement noise.

\section{Simulation-based calibration}
\label{app:sbc}

Simulation-based calibration (SBC) tests whether the nominal credible
intervals reported by the posterior estimator have the claimed frequentist
coverage~\cite{talts2018sbc}. An interval with nominal coverage $\alpha$
should contain the true parameter value in a fraction $\alpha$ of repeated
experiments. SBC operationalizes this check using synthetic test cases for
which the true parameter values are known.

\textbf{The self-consistency property.} Bayesian inference satisfies a
fundamental self-consistency condition: if a true parameter value
$\bm{\theta}^*$ is drawn from the prior $p(\bm{\theta})$ and synthetic data
$\mathbf{y}^*$ are simulated from the forward model
$p(\mathbf{y}\mid\bm{\theta}^*,t^*)$, then, for an \emph{exact} posterior,
the true value $\theta^*_j$ (for each scalar parameter component $j$) is
statistically indistinguishable from a random draw from the posterior
$p(\theta_j\mid\mathbf{y}^*,t^*)$. Equivalently, the rank of $\theta^*_j$
among $L$ posterior samples drawn from $p(\theta_j\mid\mathbf{y}^*,t^*)$
is uniformly distributed over $\{0, 1, \ldots, L\}$.

\textbf{Computation.} For each held-out test case $i=1,\ldots,N_\mathrm{test}$:
\begin{enumerate}[noitemsep]
  \item Take the true parameter vector $\bm{\theta}^{(i)}$ and thickness
    $t^{(i)}$ from the held-out test set, which sample the uniform prior
    ranges in Table~\ref{tab:fe_inputs}, and let
    $\mathbf{x}^{(i)}=(\bm{\theta}^{(i)},t^{(i)})$ denote the corresponding GP
    input.
  \item Draw a synthetic observation $\mathbf{y}^{(i)}$ from the stochastic
    observation model used for CFM training,
    $y_k^{(i)} \sim \mathcal{N}\bigl(\hat{y}_k(\mathbf{x}^{(i)}),\;
    \sigma^2_{\mathrm{GP},k}(\mathbf{x}^{(i)}) + \nu_k\bigr)$, as in
    Eq.~\eqref{eq:gp_aug}.
  \item Draw $L$ posterior samples
    $\{\hat{\bm{\theta}}^{(s)}\}_{s=1}^{L}$ from the CFM estimator
    conditioned on $(\mathbf{y}^{(i)}, t^{(i)})$.
  \item For each parameter component $j$, compute the rank
    $r_j^{(i)} = \#\bigl\{s : \hat{\theta}_j^{(s)} < \theta_j^{(i)}\bigr\}$.
\end{enumerate}
In the present study, $N_\mathrm{test}=90$ and $L=2000$.

\textbf{Rank histograms.} The ranks $\{r_j^{(i)}\}_{i=1}^{N_\mathrm{test}}$
are collected across all test cases and displayed as a histogram. For a
calibrated posterior, the rank histogram should be approximately flat
(uniform). Systematic deviations indicate specific types of miscalibration:
a U-shaped histogram (excess of very low and very high ranks) indicates
overconfidence---the posterior is too narrow and the true value frequently
falls near the tails; a mound-shaped histogram indicates
underconfidence---the posterior is too wide; a one-sided skew indicates a
systematic bias in the posterior location.

\textbf{Empirical coverage curves.} The empirical coverage at nominal level
$\alpha$ is the fraction of test cases in which the true scalar parameter
value falls inside the $\alpha$-level highest posterior density (HPD)
interval for that component:
\begin{equation}
    \widehat{\mathrm{Cov}}_j(\alpha)
    =
    \frac{1}{N_\mathrm{test}}
    \sum_{i=1}^{N_\mathrm{test}}
    \mathbf{1}\!\left[
    \theta_j^{(i)}
    \in
    \mathrm{HPD}_{\alpha,j}\!\left(\bigl\{\hat{\theta}_j^{(s)}\bigr\}_{s=1}^{L}\right)
    \right].
    \label{eq:coverage}
\end{equation}
For a calibrated posterior, the empirical coverage curve should follow the
diagonal $\widehat{\mathrm{Cov}}_j(\alpha)=\alpha$. A curve that falls
\emph{below} the diagonal indicates overconfidence (the intervals are too
narrow); a curve that falls \emph{above} the diagonal indicates
underconfidence (the intervals are too wide).

\bibliographystyle{elsarticle-num}
\bibliography{cas-refs}

\end{document}